\documentclass[11pt]{article}

\usepackage[preprint]{acl}
\usepackage{times}
\usepackage{latexsym}
\usepackage{booktabs}
\usepackage{rotating}
\usepackage[T1]{fontenc}
\usepackage{pifont}
\usepackage[most]{tcolorbox}
\usepackage{multirow}
\usepackage[table,xcdraw]{xcolor}
\usepackage{titletoc}
\usepackage{balance}

\usepackage[utf8]{inputenc}

\usepackage{microtype}

\usepackage{inconsolata}

\usepackage{graphicx}

\title{WebCoderBench: Benchmarking Web Application Generation with Comprehensive and Interpretable Evaluation Metrics}

\author{
 \textbf{Chenxu Liu\textsuperscript{1}}\thanks{Equal contribution.}\thanks{The email address of the first author is: \href{mailto:chenxuliu@stu.pku.edu.cn}{chenxuliu@stu.pku.edu.cn}},
 \textbf{Yingjie Fu\textsuperscript{1}}\footnotemark[1],
 \textbf{Wei Yang\textsuperscript{3}},
 \textbf{Ying Zhang\textsuperscript{2}\textsuperscript{4}\footnotemark[3]},
 \textbf{Tao Xie\textsuperscript{1}\textsuperscript{4}\thanks{Corresponding authors.}}
\\
 \textsuperscript{1}Key Lab of HCST (PKU), MOE; SCS, Peking University, China
 \\
 \textsuperscript{2}Key Lab of HCST (PKU) \& NERCSE, MOE; Peking University, China
 \\
 \textsuperscript{3}University of Texas at Dallas, USA
\\
\textsuperscript{4}Beijing Tongming Lake Information Technology Application Innovation Center, China
\\
 \small{
   \textbf{Correspondence:} \href{mailto:zhang.ying@pku.edu.cn}{zhang.ying@pku.edu.cn}, \href{mailto:taoxie@pku.edu.cn}{taoxie@pku.edu.cn}
 }
}

\begin{document}


\maketitle
\begin{abstract}
Web applications (web apps) have become a key arena for large language models (LLMs) to demonstrate their code generation capabilities and commercial potential. However, building a benchmark for LLM-generated web apps remains challenging due to the need for real-world user requirements, generalized evaluation metrics without relying on ground-truth implementations or test cases, and interpretable evaluation results. To address these challenges, we introduce WebCoderBench, the first real-world, generalized, and interpretable benchmark for web app generation. WebCoderBench comprises 1,572 real user requirements, covering diverse modalities and expression styles that reflect realistic user intentions. WebCoderBench provides 24 fine-grained evaluation metrics across 9 perspectives, combining the rule-based and LLM-as-a-judge paradigms for fully automated, objective, and general evaluation. Moreover, WebCoderBench adopts human-preference-aligned weights over metrics to yield interpretable overall scores.
Experiments across 12 representative LLMs and 2 LLM-based agents show that there exists no dominant model across all evaluation metrics, offering an opportunity for LLM developers to optimize their models in a targeted manner for a more powerful version.
\end{abstract}

\section{Introduction}

Web applications (web apps), leveraging their standardized, lightweight, cross-platform characteristics and vast customization ecosystem, have emerged as a key battleground for large language models (LLMs) to capitalize on their commercial potential. In industrial scenarios, LLMs take user requirements as input and automatically generate the code of the corresponding web app as output.
For these LLMs, a real-world, generalized, and interpretable benchmark not only facilitates objective understanding of their capabilities but also provides critical direction for subsequent optimization.

Benchmarking LLM-generated web apps demands substantial manual effort and specialized design to address three key challenges.
First, the dataset of the benchmark should be collected from real-world user requirements \textbf{(authenticity)}. Due to the various backgrounds, users can describe their requirements in different styles and expect web apps of diverse complexities and specialization. It is necessary to keep the authentic user requirements to reflect real usage. 
Second, the evaluation metrics of the benchmark should be general to accommodate the open-ended nature of natural-language instructions \textbf{(generality)}. Real-world requirements are often vague and can be satisfied through multiple implementations and designs, making it impractical to rely on fixed ground-truth implementations or test cases.
Third, the evaluation results produced by the benchmark should both align with user preferences and provide interpretable insights \textbf{(interpretability)}. User satisfaction with web apps is inherently multi-dimensional, shaped by diverse preferences that correspond to different LLM capabilities. Hence, fine-grained and interpretable evaluation outcomes are essential for comprehensive analysis and targeted improvement.

Existing benchmarks for evaluating LLMs on web app generation fall into three main categories, each facing limitations in addressing the aforementioned challenges.
First, synthetic benchmarks~\cite{webgenbench,fullfront,frontendbench,webbench,artifactsbench} construct datasets using LLM-generated or expert-written natural language requirements, failing to capture the diversity and authenticity of real-world user expressions.
Second, reference-based benchmarks~\cite{pix2code,design2code,sketch2code,vision2ui,web2code,webcode2m} require LLMs to reproduce web apps from provided screenshots or sketches, thus lacking the generality needed to evaluate open-ended user requirements.
Third, arena-style benchmarks~\cite{WebDevArena,designbench} collect real-world user requirements and rank LLMs through large-scale human voting in blind evaluations~\cite{arenaranking}. While these results align with human preferences, such benchmarks depend heavily on manual annotations and fail to deliver fine-grained, quantitative evaluations for a deep analysis.

To fill this gap, we propose WebCoderBench, the first real-world, generalized, and interpretable benchmark offering comprehensive and automated evaluation metrics for LLM-based web app generation. 
For \textbf{authenticity}, WebCoderBench contains 1,572 real-world user requirements collected from an online LLM service of our industrial partner. The collected requirements span multiple modalities and cover a wide range of expression styles, from precise to ambiguous and from technical to colloquial, corresponding to expected web app artifacts of varying complexity, faithfully capturing the expressions of real users.
For \textbf{generality}, WebCoderBench provides 24 fine-grained evaluation metrics across 9 perspectives, combining rule-based metrics with the LLM-as-a-judge paradigm to achieve fully automated and objective evaluation without relying on ground-truth implementations or test cases.
For \textbf{interpretability}, WebCoderBench not only reports scores of individual metrics, but also leverages user-preference-based weights across metrics to derive an overall score that aligns with real-world user priorities, providing both quantitative and human-aligned insights.

Experiments across 12 representative LLMs and 2 LLM-based agents demonstrate that WebCoderBench provides interpretable and human-aligned evaluations of web app generation. The results reveal users’ varying preferences across different perspectives of web app quality and uncover the strengths and weaknesses of existing models in detail, offering actionable insights for improvement. Notably, even the most advanced LLMs fail to achieve the highest scores across all 24 metrics.

In summary, this paper makes the following main contributions:
\begin{itemize}
    \item We present WebCoderBench, the first benchmark that enables comprehensive, interpretable, and automated evaluation for web app generation.
    \item We build a dataset of 1,572 real user requirements, faithfully capturing real-world usage.
    \item We design 24 fine-grained evaluation metrics spanning 9 perspectives, integrating the rule-based and LLM-as-a-judge paradigms, with human-preference-based weighting for interpretability and fairness.
    \item We evaluate 12 representative LLMs and 2 LLM-based agents on our benchmark. The results show that WebCoderBench can provide interpretable results for targeted improvements of existing LLMs. Our results are publicly available\footnote{\url{https://huggingface.co/spaces/CompileError/WebCoderBench}}.
\end{itemize}

\section{Related Work}
This section first introduces LLM-driven systems capable of generating web apps, and then lists benchmarks for evaluating their capabilities.

\subsection{LLM-Driven Systems for Web App Generation}
LLM-driven systems for web app generation mainly fall into three categories.
The first category is foundation models, including general-purpose models (e.g., GPT5~\cite{GPT}, Claude~\cite{Claude}, Gemini~\cite{Gemini}, DeepSeek~\cite{DeepSeek}, Qwen~\cite{Qwen3}, Doubao~\cite{Doubao}) and code-specialized models (e.g., DeepSeek-Coder~\cite{deepseekcoder}, Qwen-Coder~\cite{Qwen2.5coder}, StarCoder~\cite{starcoder}).
These models serve as the core reasoning engines, capable of multi-language code synthesis, long-context reasoning, and supporting iterative repairs  via execution feedback.
The second category is IDE/CLI-centric coding agents (e.g., Cursor~\cite{Cursor}, Trae~\cite{TRAE}, Windsurf~\cite{Windsurf}, and Claude Code~\cite{ClaudeCode}). 
These agents support manipulating local projects through project bootstrapping, dependency resolution, multi-file editing, and command-line tool control.
The third category is chat-based coding agents (e.g., Manus~\cite{Manus}, MiniMax agent~\cite{Minimax-agent}, Doubao Coding~\cite{DoubaoCoding}, and Lovable~\cite{Lovable}). These agents provide an accessible conversational interface to translate natural language requirements directly into web applications, often utilizing cloud-based sandboxes for instant rendering and deployment without requiring local environment setup.

\subsection{Benchmarks for Web App Generation}

{
\setlength{\tabcolsep}{1pt}
\begin{table*}[t]
\caption{Comparison with existing related benchmarks.}
\resizebox{\textwidth}{!}{
\begin{tabular}{ccclclc}
\toprule
\textbf{Benchmark}      & \textbf{Modal}          & \textbf{Sample} & \multicolumn{1}{c}{\textbf{Source}} & \textbf{Automated} & \multicolumn{1}{c}{\textbf{Metrics}}                  & \textbf{Ground Truth} \\ \midrule
WebGen-Bench~\cite{webgenbench}   & Text           & 101    & LLM + human experts          & Yes                  & LLM as a judge + test case execution           & Test cases   \\
FullFront~\cite{fullfront}   & Text + Image           & 1,800    & LLM          & Yes                  & Code \& visual similarity + LLM as a judge           & Code + Images   \\
FrontendBench~\cite{frontendbench}  & Text           & 148    & LLM                        & Yes                  & Test case execution                          & Test cases   \\
Web-Bench~\cite{webbench}      & Text           & 100    & Human experts              & Yes                  & Test case execution                          & Test cases   \\
WebDev Arena~\cite{WebDevArena}    & Text           & 10,501  & Real user requirements     & No                   & Pairwise manual labeling                             & -            \\
Design Arena~\cite{DesignArena}   & Text           & -      & Real user requirements     & No                   & Pairwise manual labeling                             & -            \\
ArtifactsBench~\cite{artifactsbench} & Text           & 1,825    & LLM + human experts          & Yes                  & LLM as a judge                               & Checklist    \\
WebCoderBench (ours)           & Text + Image + URL & 1,572    & Real user requirements     & Yes                  & (Rule-based + LLM as a judge) * preference & Checklist    \\ \bottomrule
\end{tabular}}
\label{related work table}
\end{table*}}

Web app generation has long been an active research topic. Since the pioneering work of pix2code~\cite{pix2code}, a large body of benchmarks and datasets~\cite{websight,design2code,sketch2code,vision2ui,web2code,webcode2m} has focused on generating web apps from screenshots or sketches. These approaches typically conduct evaluation by comparing the generated web app code or rendered screenshots against the corresponding ground-truth code or original screenshots.

In recent years, the emergence of LLMs has greatly lowered the barrier to web app generation, enabling users to create customized applications through natural-language or multi-modal requirements.
However, as summarized in Table~\ref{related work table}, existing benchmarks fall short of addressing the full spectrum of the three challenges in this domain.
First, some benchmarks rely on LLM-generated or expert-curated requirements, failing to capture the authenticity and diversity of real user behavior.
Second, some benchmarks depend on predefined test cases or ground-truth code for evaluation.
Third, some benchmarks are not fully automated, heavily relying on manual labeling.

\section{Dataset}

WebCoderBench includes 1,572 real-world user requirements across 20 application categories. Of these requirements, 1,413 are text-only, 123 include images, and 36 include URLs as additional resources. In terms of clarity, 78 requirements are vague, 730 are intermediate, and 764 are clear. We further divide these 1,572 requirements into five complexity levels with 179, 433, 658, 259, and 43 samples from simple to complex, respectively. Overall, the dataset exhibits sufficient diversity to represent real-world user requirements.

\subsection{Data Collection}
\label{sec:datacollection}

\begin{figure}[t]
\centerline{\includegraphics[width = \columnwidth]{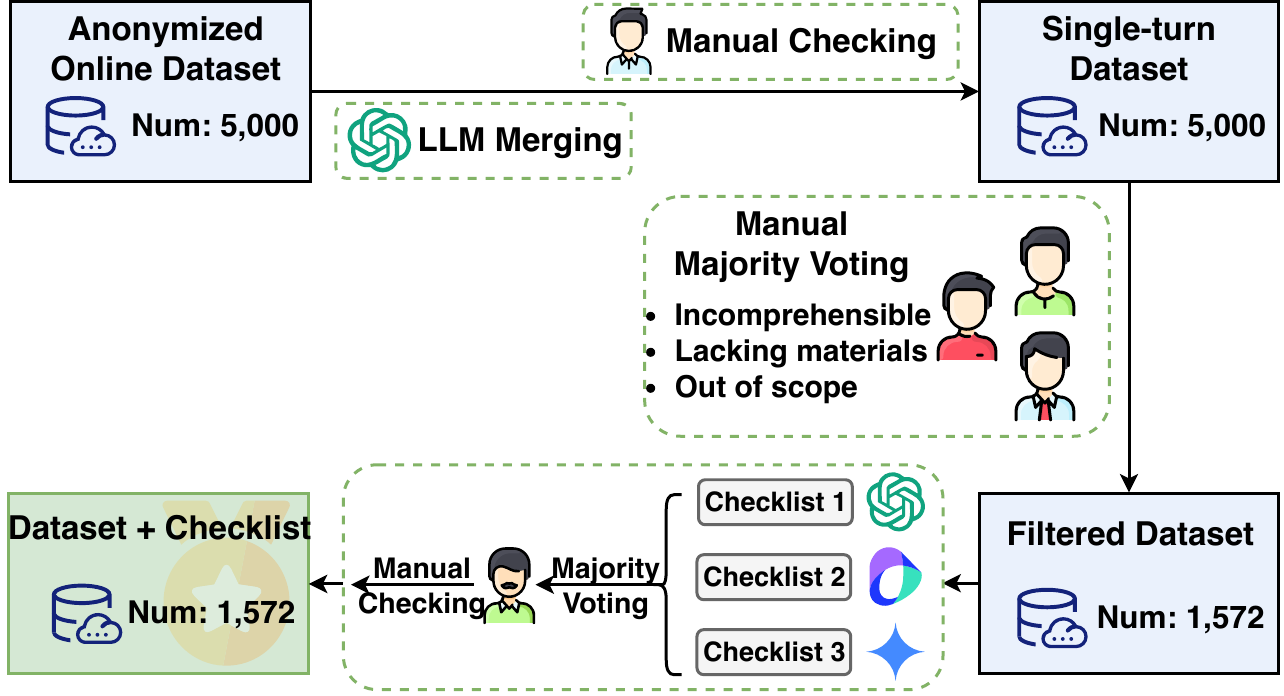}}
\caption{The dataset construction process of WebCoderBench.}
\label{dataset construction}
\end{figure}

The data collection pipeline of WebCoderBench is illustrated in Figure~\ref{dataset construction}. We construct the original dataset by collecting one week of anonymized and filtered real-world online data from our industrial partner and randomly sampling 5,000 user requirements.

First, to reduce bias caused by revision turns that depend on specific model outputs in multi-turn requirements, we merge multi-turn requirements into single-turn ones by human experts assisted by an LLM (Gemini-2.5-pro). Manual revisions are made to ensure fluency and content anonymization.

Next, three human experts review each requirement to remove those that are incomprehensible, lack supplementary materials (e.g., images), or are inapplicable to native HTML scenarios. We further conduct text-level and semantic-level de-duplication using MinHash~\cite{MinHash} and MiniLM~\cite{MiniLM} semantic embeddings, following the practice of ArtifactsBench~\cite{artifactsbench}. The filtered dataset retains 1,572 requirements.

\begin{figure}[t]
\centerline{\includegraphics[width = \columnwidth]{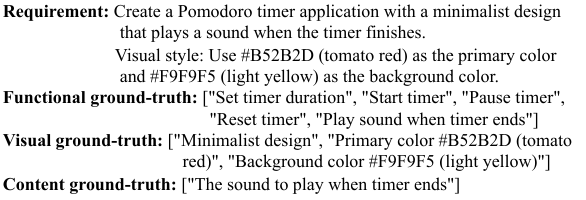}}
\caption{An example user requirement with its corresponding ground-truth checklists.}
\label{GT example}
\end{figure}

Finally, to enable objective evaluation, we establish ground-truth checklists for each requirement across three dimensions: functionality, visual design, and content. We adopt three LLMs (GPT-5-Chat-2025-08-07, Gemini-2.5-pro, and Doubao-Seed-1.6) to infer ground-truth checklists for each dimension. After that, human experts merge and validate the outputs to produce the final ground-truth checklists. Figure~\ref{GT example} shows an example user requirement with its corresponding ground-truth checklists. The checklists comprise only high-level and minimal requirement points that the web app artifact should satisfy.

\subsection{Dataset Statistics}


We perform detailed and multi-dimensional classifications of each requirement to enable an in-depth analysis of the characteristics and distribution of the dataset. The results are shown in Table~\ref{dataset detail table}, with all tags manually labeled by human experts assisted by an LLM (Gemini-2.5-pro). 

{
\setlength{\tabcolsep}{2pt}
\begin{table}[t]
\caption{Dataset statistics of WebCoderBench.}
\resizebox{\columnwidth}{!}{
\begin{tabular}{@{}lrlr@{}}
\toprule
\textbf{Type}           & \textbf{Number} & \textbf{Type}             & \textbf{Number} \\ \midrule
\textbf{Application Category} &                 & \textbf{Clarity of Requirement} &                 \\
- AI-powered                  & 74              & - Clear                         & 764             \\
- BBS                         & 6               & - Intermediate                  & 730             \\
- Corporate   Website         & 41              & - Vague                         & 78              \\
- Data Visualization          & 59              & \textbf{Style of Expression}    &                 \\
- E-commerce                  & 40              & - Technical                     & 683             \\
- Enterprise Backend          & 116             & - Colloquial                    & 724             \\
- Entertainment               & 435             & - Role-playing                  & 60              \\
- Fintech                     & 32              & - Analogy                       & 105             \\
- Health   Care               & 14              & \textbf{Artifact Complexity}    &                 \\
- IoT Interface               & 10              & - Highly Simple                 & 179             \\
- Multimedia                  & 34              & - Simple                        & 433             \\
- News Media                  & 5               & - Medium                        & 658             \\
- Online Education            & 131             & - Complex                       & 259             \\
- Online Office Suite         & 3               & - Highly Complex                & 43              \\
- Personal Webite             & 49              & \textbf{Input Modality}            &                 \\
- Public Service              & 27              & - Text Only                     & 1,413           \\
- Scientific Demo             & 69              & - Text with Images              & 123             \\
- Social Media                & 13              & - Text with URLs                & 36              \\
- Tourism                     & 10              &                                 &                 \\
- Utility Website             & 404             &                                 &                 \\ \bottomrule
\end{tabular}}
\label{dataset detail table}
\end{table}
}

In terms of application category, WebCoderBench covers a wide range of web apps, including Utility Websites (404 requirements), Entertainment (435), and Online Education (131), which together account for the majority of the samples, reflecting the diversity of real-world web scenarios. Less frequent but important categories, such as AI-powered, Fintech, and Scientific Demo, are also included, showing the variety of the dataset.

Regarding clarity, most requirements are clear (764) or intermediate (730), while only a small portion (78) are vague, showing that our dataset primarily focuses on interpretable and executable tasks, but also retains challenging tasks for LLMs to explore. The style of expression dimension indicates that both technical (683) and colloquial (724) descriptions are prevalent, capturing different levels of formality in user expression, with occasional role-playing (60) and analogy-based (105) requirements adding linguistic diversity.

The artifact complexity dimension spans from highly simple to highly complex outputs, with most samples falling into medium (658) and simple (433) levels, showing balanced difficulty for model evaluation. In terms of input modality, our dataset mainly includes text-only requirements (1,413), complemented by text with images (123) and text with URLs (36), allowing evaluation of both uni-modal and multi-modal understanding abilities. 
It is worth noting that, unlike the task of generating code from screenshots~\cite{pix2code,webcode2m}, our user requirements can include images and URLs that are intended to serve as page content rather than as reference designs.

In terms of length statistics, the shortest requirement in our dataset contains 3 characters, while the longest contains 16,198 characters, with an average length of 313.74 characters. Measured using the GPT-5~\cite{GPT} tokenizer, the shortest requirement contains 3 tokens, the longest 9,235 tokens, and the average length is 202.31 tokens.

In summary, these statistics show that our dataset covers a wide variety of web app types, styles, and complexities, making it promising for evaluating LLMs in web app generation.

\section{Evaluation Metrics}

To comprehensively assess the quality of web app artifacts generated by LLMs, we design a set of evaluation metrics from multiple perspectives. Inspired by practices in the text-to-image domain~\cite{text2imagesurvey}, our evaluation considers two major aspects: general quality and alignment quality. These aspects are further divided into nine perspectives encompassing 24 detailed evaluation metrics, jointly providing a comprehensive assessment. 
The metrics are defined based on public standards (e.g., Google Lighthouse~\cite{Lighthouse}, Lint~\cite{HTMLHint,StyleLint,ESLint}, W3C design principles~\cite{W3C}) and internal guidelines from our industrial partner (e.g., visual design and copywriting standards). All the metrics are listed in Table~\ref{metrics table}, with detailed information provided in Appendix~\ref{Detailed Evaluation Metrics}. Each metric produces a quantitative score ranging from 0 to 100, where a higher score indicates higher quality.

{\setlength{\tabcolsep}{1pt}
\begin{table}[t]
\caption{The evaluation metrics used in WebCoderBench, with rule-based metrics in white background, while metrics using the LLM-as-a-judge paradigm are shaded in gray. The ``Input'' and ``Render'' columns indicate the input modality (S: screenshot; C: code) of each evaluation metric and whether it requires page rendering.}
\resizebox{\columnwidth}{!}{
\begin{tabular}{@{}cllcc@{}}
\toprule
\textbf{Aspects}                                                              & \multicolumn{1}{c}{\textbf{Perspectives}} & \multicolumn{1}{c}{\textbf{ID + Evaluation Metrics}}           & \textbf{Input} & \textbf{Render} \\ \midrule
                                                                              &                                           & \cellcolor[HTML]{EFEFEF}1. General Functionality Correctness & C           & \ding{55}           \\
                                                                              &                                           & 2. Best Practices                                            & C           & \ding{51}            \\ & &
 3. Error Handling                                            & C           & \ding{55} \\          
                                                                              &                                           & 4. Runtime Console Errors                                    & C           & \ding{51}            \\
                                                                              & \multirow{-5}{*}{Code Quality}            & 5. Static Syntax Checking (Linting)                          & C           & \ding{55}           \\ \cmidrule(){2-5} 
                                                                              &                                           & \cellcolor[HTML]{EFEFEF}6. General Visual Experience         & S     & \ding{51}            \\
                                                                              &                                           & 7. Component Style Consistency                               & C           & \ding{55}           \\
                                                                              &                                           & 8. Icon Style Consistency                                    & C           & \ding{51}           \\
                                                                              &                                           & 9. Layout Consistency                              & S          & \ding{51}            \\
                                                                              &                                           & 10. Layout Sparsity                              & S          & \ding{51}            \\
                                                                              & \multirow{-6}{*}{Visual Quality}          & 11. Visual Harmony Degree                                     & S     & \ding{51}            \\ \cmidrule(){2-5}
                                                                              &                                           & 12. Copywriting Quality                                       & C           & \ding{55}           \\
                                                                              &                                           & 13. Media Quality                                             & C           & \ding{55}           \\
                                                                              &                                           & 14. Placeholder Quality                                       & C           & \ding{55}           \\
                                                                              & \multirow{-4}{*}{Content Quality}         & 15. Resource Validity                                         & C           & \ding{55}           \\ \cmidrule(){2-5} 
                                                                              & Performance Quality                       & 16. General Performance                                       & C           & \ding{51}            \\ \cmidrule(){2-5} 
                                                                              &                                           & 17. Accessibility Core Metrics                                & C           & \ding{51}            \\
                                                                              &                                           & 18. Cross-Browser Compatibility                               & C           & \ding{51}            \\
                                                                              & \multirow{-3}{*}{Accessibility}           & 19. Mobile Device Compatibility                               & C           & \ding{51}            \\ \cmidrule(){2-5} 
                                                                              &                                           & 20. Code Redundancy Rate                                      & C           & \ding{51}            \\
\multirow{-21}{*}{\begin{tabular}[c]{@{}c@{}}General\\ Quality\end{tabular}}  & \multirow{-2}{*}{Maintainability}         & 21. Comment Rate                                              & C           & \ding{55}           \\ \midrule
                                                                              & Functional Alignment                      & \cellcolor[HTML]{EFEFEF}22. Functional Alignment              & C           & \ding{55}           \\ \cmidrule(){2-5} 
                                                                              & Visual Alignment                          & \cellcolor[HTML]{EFEFEF}23. Visual Alignment                  & C           & \ding{55}           \\ \cmidrule(){2-5} 
\multirow{-3}{*}{\begin{tabular}[c]{@{}c@{}}Alignment\\ Quality\end{tabular}} & Content Alignment                         & \cellcolor[HTML]{EFEFEF}24. Content Alignment                 & C           & \ding{55}           \\ \bottomrule
\end{tabular}}
\label{metrics table}
\end{table}
}

\subsection{General Quality}
General quality measures the overall quality of the generated web app, regardless of the specific user requirement. We evaluate general quality from six distinct perspectives, using a combination of automated rule-based metrics and the LLM-as-a-judge paradigm (with Gemini-2.5-pro as the judge according to our experience).

\textbf{Code Quality} evaluates the functional correctness and implementation of the generated code, focusing on correctness, robustness, and adherence to engineering practices.

\textbf{Visual Quality} evaluates the visual presentation and design of the generated web app, emphasizing layout consistency, stylistic coherence, and overall aesthetic experience.

\textbf{Content Quality} evaluates the informativeness and the quality of resources loaded in the generated web app.

\textbf{Performance Quality} evaluates the runtime behavior of the generated web app, measuring the efficiency of page rendering and resource loading.

\textbf{Accessibility} evaluates the disability-friendliness and platform compatibility of the generated web app, assessing usability across different devices and browsers.

\textbf{Maintainability} evaluates the readability, reusability, and ease of long-term maintenance of the generated code.

\subsection{Alignment Quality}
Alignment quality measures how well the generated web app meets the corresponding user requirements. We evaluate alignment quality using human-labeled ground-truth checklists (Section~\ref{sec:datacollection}) as references and employing the LLM-as-a-judge paradigm (with GPT-5-chat as the judge according to our experience). 
The three metrics, namely \textbf{Functional Alignment}, \textbf{Visual Alignment}, and \textbf{Content Alignment}, evaluate the consistency between the generated web app and the user requirement in terms of functionalities, visual appearance, and textual/multimedia content, respectively.

\subsection{Weight Assignment}

\begin{figure}[t]
\centerline{\includegraphics[width = \columnwidth]{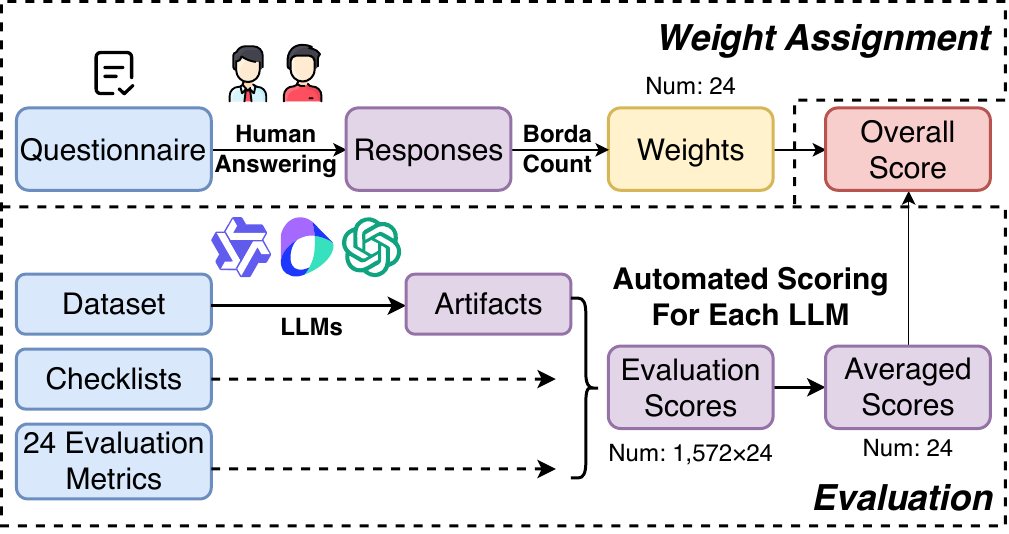}}
\caption{The weight assignment and evaluation workflow of WebCoderBench.}
\label{evaluation_process}
\end{figure}

After defining evaluation metrics, a crucial process is to combine the outcome scores in a meaningful way to generate an overall score, reflecting human preference. Our process of generating the overall score is shown in Figure~\ref{evaluation_process}. Existing studies~\cite{webuibench,artifactsbench} typically assign weights to each metric either uniformly or heuristically. However, real-world users do not treat all the metrics equally and can prioritize certain perspectives (e.g., emphasizing code quality while paying less attention to maintainability). Consequently, heuristic weighting neither aligns with user priorities nor ensures fairness.

To obtain preference-aligned weights for each metric, we conduct an internal survey with our industrial partner. Specifically, we ask each participant to rank the 24 metrics according to their perceived importance. To reduce the cognitive burden on participants, each of them is first asked to rank the nine perspectives listed in Table~\ref{metrics table}, and subsequently rank the metrics within each perspective. During a three-day survey period, our questionnaire receives 1,076 views and 899 responses, yielding a response rate of 83.55\%. We further filter the responses by completion time, retaining only those that take more than two minutes to complete. As a result, we obtain 141 valid responses. According to a recent study~\cite{surveysample}, a sample size of more than 100 responses can be considered as sufficient for ranking and regression analyses. Among the valid responses, there are 21 data scientists, 19 product managers, 22 legal personnel, 23 front-end/back-end developers, 2 designers, 39 operations staff, and 15 quality assurance personnel, according to their user personas.

\begin{figure}[t]
\centerline{\includegraphics[width = \columnwidth]{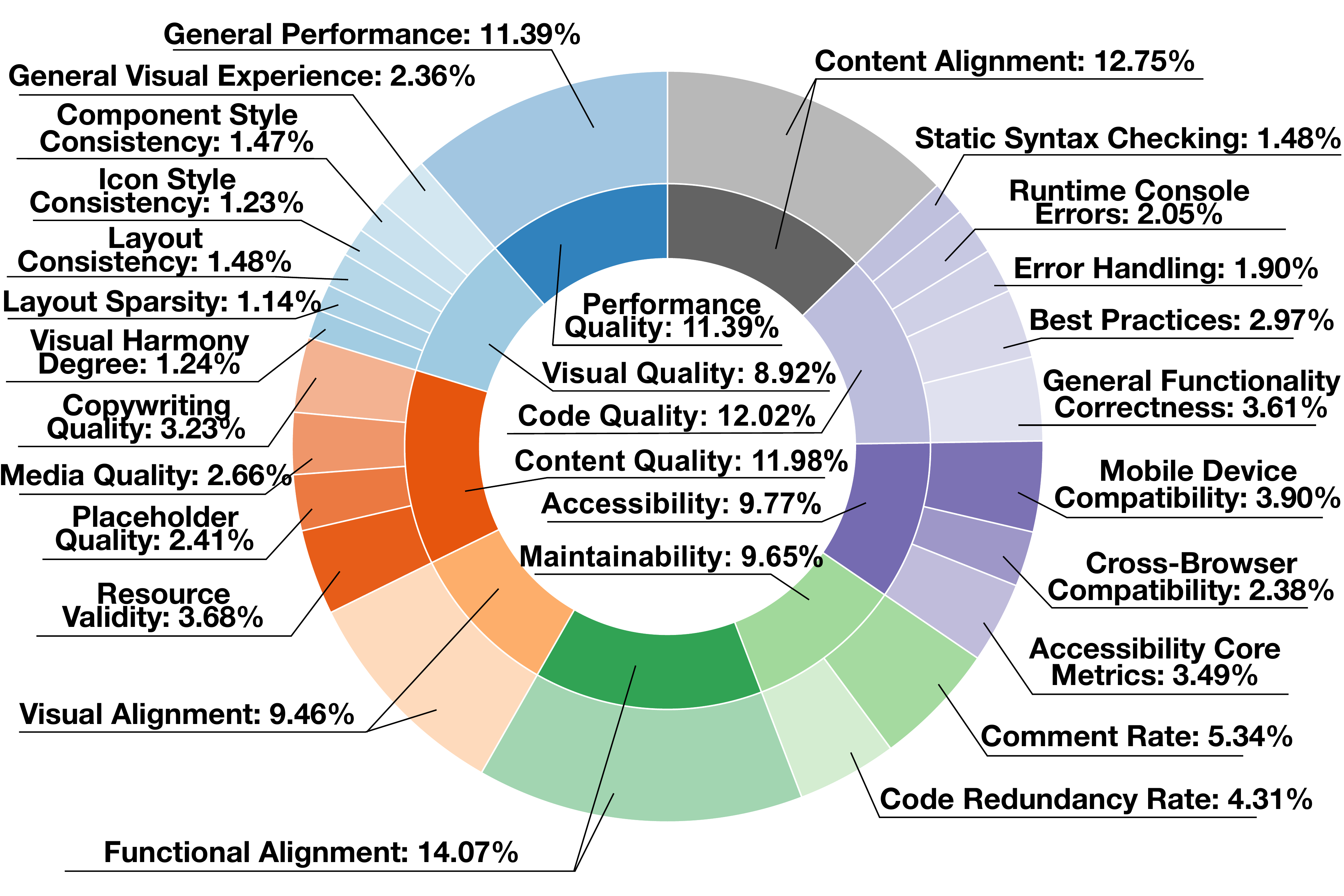}}
\caption{The weight proportion of each perspective and each evaluation metric.}
\label{weight pie chart}
\end{figure}

We further apply the Borda Count~\cite{borda} algorithm to extract weights from responses. This algorithm assigns a score to each item (perspective or metric) based on its position in each participant’s ranking. The weight of each item is then obtained by summing its scores across all participants and normalizing by the total score over all items. The weight of each metric is calculated by multiplying its own weight by the weight of the perspective to which it belongs. The weights of all the metrics sum to 1. The resulting weights are shown in Figure~\ref{weight pie chart}.

The overall score of each model is calculated by summing the z-scores~\cite{zscore} of all the metrics, averaged over the 1,572 samples in the dataset, and weighted by the corresponding metric weights. We adopt z-scores to ensure that the scores of different metrics are on a comparable scale with consistent discriminability, thereby emphasizing the effect of the weights. The z-scores are calculated by $z_{i,j} = \frac{x_{i,j} - \mu_j}{\sigma_j}$, where $x_{i,j}$ denotes the raw score of the $i$-th sample on the $j$-th metric, and $\mu_j$ and $\sigma_j$ denote the mean and standard deviations of this metric over all samples, respectively.

Through this process, we establish a data-driven mechanism that grounds the metric weighting scheme in authentic human preferences rather than arbitrary heuristics. The derived weights effectively capture how users implicitly trade off among different quality perspectives, enabling our overall evaluation score to reflect real-world user preferences.





\section{Evaluation Results and Analysis}

\subsection{Settings}

{\setlength{\tabcolsep}{1pt}
\begin{table*}[t]
\caption{Weighted z-scores for each perspective, the overall score, and the ranking of each model. The percentage value indicates how many standard deviations the weighted value deviates from the average. Rows 1–8 correspond to open-source LLMs, rows 9–12 to closed-source LLMs, and rows 13–14 to LLM-based agents. Note that the scores of LLMs are averaged among the whole dataset, while those of LLM-based agents are averaged among a random subset of 165 requirements.}
\resizebox{\textwidth}{!}{
\begin{tabular}{clrrrrrrrrrrr}
\hline
\textbf{ID} & \textbf{Model}           & \textbf{\begin{tabular}[c]{@{}c@{}}Code\\ Quality\end{tabular}} & \textbf{\begin{tabular}[c]{@{}c@{}}Visual\\ Quality\end{tabular}} & \textbf{\begin{tabular}[c]{@{}c@{}}Content\\ Quality\end{tabular}} & \textbf{\begin{tabular}[c]{@{}c@{}}Performance\\ Quality\end{tabular}} & \textbf{Accessibility} & \textbf{Maintainability} & \textbf{\begin{tabular}[c]{@{}c@{}}Functional\\ Alignment\end{tabular}} & \textbf{\begin{tabular}[c]{@{}c@{}}Visual\\ Alignment\end{tabular}} & \textbf{\begin{tabular}[c]{@{}c@{}}Content\\ Alignment\end{tabular}} & \textbf{\begin{tabular}[c]{@{}c@{}}Overall\\ Score (Sum)\end{tabular}} & \textbf{Ranking} \\ \hline
1           & DeepSeek-R1-0528         & -0.31\%                                                         & -1.27\%                                                           & 1.81\%                                                             & 0.28\%                                                                 & 1.33\%                 & 1.09\%                   & -0.98\%                                                                 & 0.21\%                                                              & 0.25\%                                                               & 2.41\%                                                                 & 7                \\
2           & DeepSeek-V3.1            & -0.86\%                                                         & -1.60\%                                                           & -1.04\%                                                            & 0.82\%                                                                 & -1.18\%                & -0.04\%                  & 0.00\%                                                                  & 0.35\%                                                              & -0.32\%                                                              & -3.88\%                                                                & 13               \\
3           & DeepSeek-V3.1-Terminus   & 0.17\%                                                          & 2.11\%                                                            & -1.58\%                                                            & 1.37\%                                                                 & 1.09\%                 & 2.58\%                   & 0.52\%                                                                  & -0.03\%                                                             & 0.03\%                                                               & 6.26\%                                                                 & 4                \\
4           & DeepSeek-V3.1-Thinking   & -0.98\%                                                         & -1.51\%                                                           & -0.44\%                                                            & 1.17\%                                                                 & -1.05\%                & -0.04\%                  & -0.14\%                                                                 & -0.24\%                                                             & -0.50\%                                                              & -3.73\%                                                                & 12               \\
5           & GLM-4.5                  & 0.28\%                                                          & 2.33\%                                                            & 5.03\%                                                             & 0.46\%                                                                 & 0.17\%                 & 0.61\%                   & 1.58\%                                                                  & 0.84\%                                                              & 0.46\%                                                               & 11.76\%                                                                & 2                \\
6           & Qwen3-Coder-Plus         & -0.63\%                                                         & -1.65\%                                                           & -0.43\%                                                            & 1.70\%                                                                 & 0.89\%                 & 1.51\%                   & -1.55\%                                                                 & -1.04\%                                                             & -1.10\%                                                              & -2.31\%                                                                & 11               \\
7           & Qwen3-235B-A22B-Instruct & -1.38\%                                                         & 0.32\%                                                            & -0.60\%                                                            & 0.68\%                                                                 & 0.22\%                 & 0.53\%                   & -0.05\%                                                                 & -0.01\%                                                             & 0.25\%                                                               & -0.04\%                                                                & 10               \\
8           & MiniMax-M2               & 1.13\%                                                          & -0.94\%                                                           & -0.56\%                                                            & 0.40\%                                                                 & -1.06\%                & 2.27\%                   & 3.20\%                                                                  & 1.50\%                                                              & 1.96\%                                                               & 7.89\%                                                                 & 3                \\ \hline
9           & Gemini-2.5-Pro           & 0.89\%                                                          & 0.99\%                                                            & -0.26\%                                                            & -2.25\%                                                                & 0.33\%                 & -1.74\%                  & 0.91\%                                                                  & 1.02\%                                                              & 0.38\%                                                               & 0.27\%                                                                 & 8                \\
10          & GPT-4o-2024-11-20        & -2.00\%                                                         & -1.69\%                                                           & -1.91\%                                                            & -2.49\%                                                                & -0.41\%                & -5.37\%                  & -11.04\%                                                                & -6.62\%                                                             & -7.12\%                                                              & -38.65\%                                                               & 14               \\
11          & GPT-5-Codex-High         & 0.41\%                                                          & 2.30\%                                                            & -1.14\%                                                            & -2.14\%                                                                & 1.78\%                 & -7.52\%                  & 2.10\%                                                                  & 1.80\%                                                              & 2.39\%                                                               & -0.03\%                                                                & 9                \\
12          & GPT-5-High               & 2.15\%                                                          & 1.36\%                                                            & -1.17\%                                                            & 1.27\%                                                                 & -2.05\%                & 0.88\%                   & 5.51\%                                                                  & 2.36\%                                                              & 3.50\%                                                               & 13.81\%                                                                & 1                \\ \hline
13          & Manus                    & 0.25\%                                                          & -0.76\%                                                           & 0.85\%                                                             & -0.56\%                                                                & -1.00\%                & 1.58\%                   & 3.29\%                                                                  & 0.33\%                                                              & 1.26\%                                                               & 5.24\%                                                                 & 5                \\
14          & MiniMax Agent            & -0.21\%                                                         & -2.45\%                                                           & 6.27\%                                                             & -1.39\%                                                                & -1.30\%                & 1.00\%                   & 1.71\%                                                                  & 0.77\%                                                              & 0.03\%                                                               & 4.43\%                                                                 & 6                \\ \hline
\end{tabular}}
\label{Main results table}
\end{table*}
}

We evaluate 12 representative LLMs and 2 LLM-based agents on WebCoderBench. The selected models span multiple families (e.g., DeepSeek~\cite{DeepSeek}, Qwen~\cite{Qwen3}, Gemini~\cite{Gemini}, GLM~\cite{GLM}, GPT~\cite{GPT}), multiple versions (e.g., base, instruct, thinking, coder), multiple modalities (e.g., text-only, multi-modal), and multiple capability types (e.g., standard~\cite{Minimax-m1,Minimax-m2}, agentic~\cite{Manus,Minimax-agent}). We invoke each LLM using its standard APIs with the recommended parameters and identical system prompts. We let each model generate a single artifact for each requirement.

For LLM-based agents, we manually enter prompts and requirements through their web interfaces to obtain the generated artifacts, and further constrain their outputs to native HTML by appending specific control prompts. Due to labor constraints, we collect artifacts for 165 requirements from these agents, corresponding to approximately one-tenth of the full dataset.

To enable uni-modal LLMs (e.g., DeepSeek) to handle multi-modal requirements, we use Gemini-2.5-pro to generate textual descriptions for each image and provide these descriptions as input to the uni-modal LLMs. The descriptions are restricted to objective visual content and do not introduce any additional information beyond what is observable in the original images. Our goal is to evaluate each model based on the modalities supports by it, rather than penalizing uni-modal models for lacking visual capabilities (e.g., assigning a score of zero or excluding them from multi-modal requirements). We believe that this setup provides a fair comparison under modality constraints.

Due to the diversity of user requirements, we observe two distinct usage patterns for images and URLs. A majority of the requirements ask the LLM to use images and URLs as reference examples for implementation (95/123 for images and 29/36 for URLs). The remaining requirements ask the LLM to include images and URLs as assets in the final generated web apps (28/123 for images and 7/36 for URLs). For all requirements, we encode images in the base64 format before passing them to the model, and URLs are included as strings within the input specification. For requirements that treat images and URLs as assets, the generated web app includes images in the base64 format and incorporates URLs as hyperlinks in the final output.

\subsection{RQ1: Main Results}

We present the weighted z-scores for each perspective, the overall scores, and the model rankings in Table~\ref{Main results table}, and the raw scores of the 24 evaluation metrics in Figure~\ref{heatmap}. Detailed z-scores for each metric are provided in Appendix~\ref{Detailed Evaluation Results} due to space limits.

\begin{figure}[t]
\centerline{\includegraphics[width = \columnwidth]{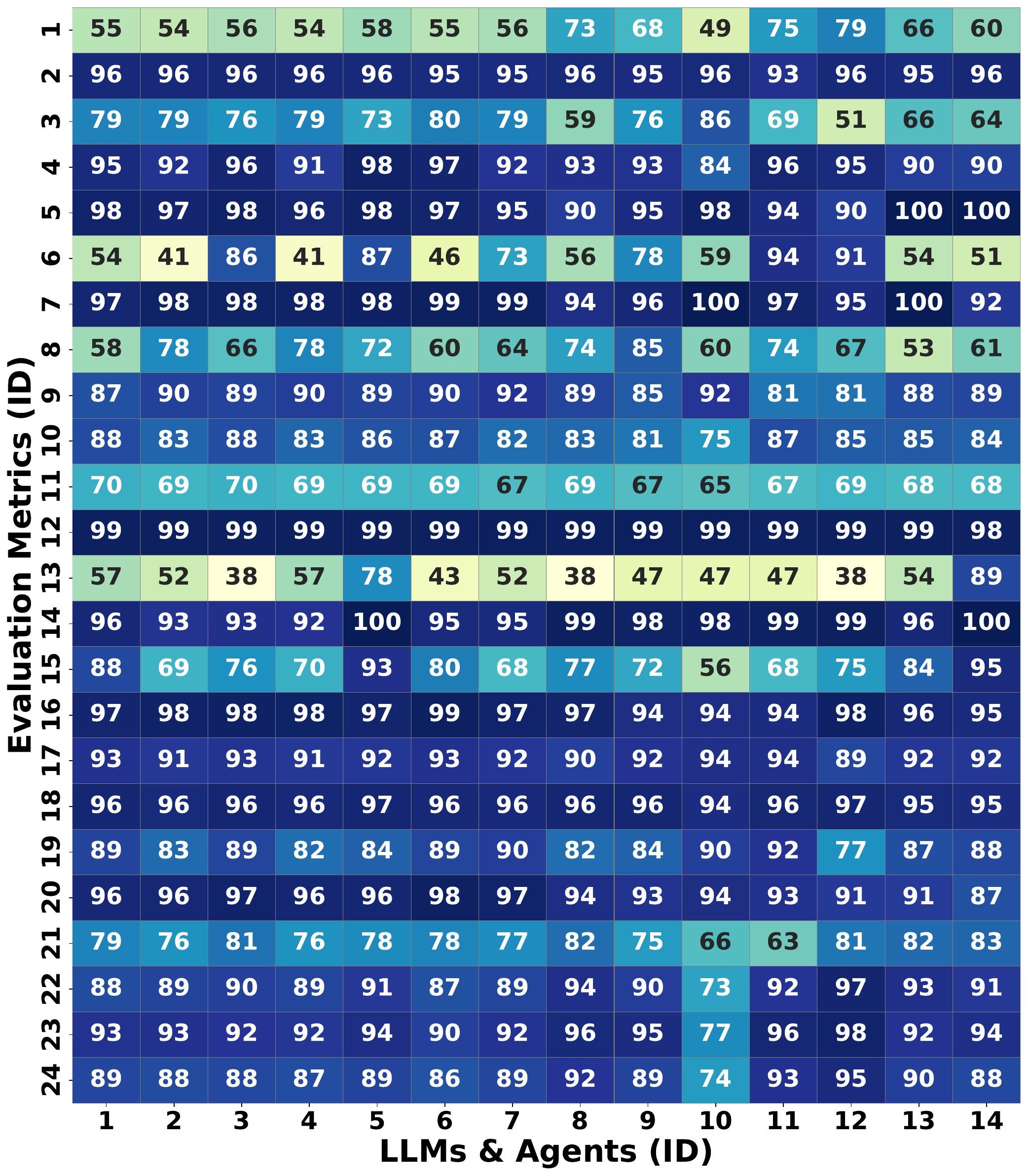}}
\caption{The detailed raw scores of 24 evaluation metrics for each LLM and LLM-based agent, with the x-axis indices denoting the IDs of evaluation metrics (corresponding to Table~\ref{metrics table}), and the y-axis indices denoting the IDs of models (corresponding to Table~\ref{Main results table}).}
\label{heatmap}
\end{figure}

Across all the evaluated LLMs and agents, GPT‑5‑High attains the highest overall score, greatly outperforming other models through its strong ability to align with requirements and its consistently positive results across most perspectives. Among the open‑source LLMs, GLM‑4.5 achieves the best effectiveness and ranks second overall, while MiniMax‑M2 also performs competitively. In contrast, GPT‑4o exhibits the weakest effectiveness, with all scores falling below the average and six out of nine perspectives ranking last among all LLMs.

There exists no single model that dominates all perspectives. GLM‑4.5 is the only model achieving above‑average effectiveness across all nine perspectives, indicating balanced and reliable capabilities. However, even GLM‑4.5 cannot outperform the average in all the 24 fine‑grained evaluation metrics, suggesting that current LLMs remain specialized rather than universally strong. The competition among the LLMs remains tight, with no single model emerging as decisively superior. Given our comprehensive and interpretable evaluation metrics, LLM developers are able to optimize their models in a targeted manner for a more powerful future version.

The comparison between open‑source and closed‑source models reveals a rapidly narrowing effectiveness gap. Although GPT‑5‑High remains the most effective model overall, the strong effectiveness of GLM‑4.5 and MiniMax‑M2 shows that recent open‑source LLMs are increasingly competitive, with less than a 6\% gap from GPT‑5‑High. 

The results also illustrate the accelerated pace of model evolution and development. The models released within the past year consistently obtain high scores across multiple evaluation perspectives, while earlier models such as GPT‑4o exhibit substantial gaps, performing below average in all perspectives and falling behind newer models by a considerable margin. This divergence highlights the rapid turnover in LLM effectiveness.

The LLM‑based agents generally score above average but fall short of expectations. Their ability to access online resources and generate complex pages results in low performance and accessibility due to the increased complexity of external resources. The complexity of the generated UI further degreades visual quality. However, these agents typically align well with user requirements due to their planning and task‑oriented reasoning capabilities. This trend can also cause an LLM-based agent to be less effective than its base model. For example, the overall score of MiniMax Agent is lower than that of MiniMax‑M2, indicating that while agents introduce enhanced capabilities, these strengths come at the cost of performance and visual consistency.

Compared to coder models (e.g., Qwen3-Coder-Plus and GPT-5-Codex), generalist models (e.g., Qwen3-Instruct and GPT-5-High) present better effectiveness, showing that the task of web app generation requires not only coding ability, but also general understanding of requirements and real-world knowledge. This finding also aligns to ArtifactsBench~\cite{artifactsbench}.

Although the weighted z-scores appear numerically close (most within 3\%), they represent differences in standard deviations from the mean for each metric, and each of them is weighted by a small percentage (shown in Figure~\ref{weight pie chart}). Therefore, even small value differences can be meaningful.

\subsection{RQ2: Effect of Question Types}

\begin{table}[t]
\caption{The overall scores averaged over the models and the questions for each question type.}
\resizebox{\columnwidth}{!}{
\begin{tabular}{lrlr}
\hline
\textbf{Type}                   & \textbf{Score} & \textbf{Type}                & \textbf{Score} \\ \hline
\textbf{Clarity of Requirement} &                & \textbf{Input Modality}      &       \\
- Clear                         &   -0.02\%             & - Text Only                  &     0.85\%           \\
- Intermediate                  &   -0.55\%             & - Text with Images           &     -8.68\%           \\
- Vague                         &   5.98\%             & - Text with URLs             &     1.54\%           \\
\textbf{Artifact Complexity}    &                & \textbf{Style of Expression} &       \\
- Highly Simple                 &   0.94\%             & - Technical                  &    -0.72\%            \\
- Simple                        &   -1.11\%             & - Colloquial                 &   1.14\%             \\
- Medium                        &   0.18\%             & - Role-playing               &    -0.89\%            \\
- Complex                       &   0.53\%             & - Analogy                    &    -1.75\%            \\
- Highly Complex                &   4.82\%             &                              &                \\ \hline
\end{tabular}}
\label{Question type table}
\end{table}

The overall scores averaged over the models and the questions for each question type are shown in Table~\ref{Question type table}. We aggregate the scores of all the models to show a general and overall trend, instead of analyzing each model separately.

We find that the results do not entirely conform to the straightforward intuition that the models tend to perform worse on vague and highly complex requirements. We manually inspect the outcomes for reasoning. For requirement clarity, the models generally perform better under vague requirements, because such descriptions provide greater freedom and reduce penalties from fine‑grained mismatches with the requirements, whereas clear requirements impose strict constraints that are easy to violate. Regarding artifact complexity, requirements of complex artifacts offer rich functional and contextual cues that help the models infer page structure (with only 3 out of 43 highly complex requirements are classified as vague), while requirements of simple artifacts lack sufficient information and thus lead to deviations. In terms of input modality, inputs with images degrade effectiveness, indicating that the models remain unstable in mapping visual content to page structures. For expression style, the models are most effective when processing colloquial requirements, indicating that the models are more suitable for the common expressions used by the general public.

\subsection{RQ3: Effect of User Personas}

The scores provided in Table~\ref{Main results table} are calculated using weights derived from all the valid questionnaire responses. However, users who belong to different user personas can prefer different perspectives. Table~\ref{User persona table} shows the weights derived from the different user personas in our collected responses.

The results indicate that the different personas exhibit preferences  that align well with the intuitive understanding over perspectives. For instance, the designers show a pronounced preference for visual quality. The code-related personas (the first four columns) tend to place greater emphasis on code quality and maintainability, whereas the operations staff and the legal personnel prioritize content quality and performance quality. Across the three alignment-related perspectives, all the personas assign high preference consistently, with the product managers and the designers exhibiting particularly strong preferences.
These patterns highlight the domain-specific expectations.

{\setlength{\tabcolsep}{3pt}
\begin{table}[t]
\caption{Weights derived from the different user personas, where each column corresponds to front-end/back-end developers (Dev), quality assurance personnel (QA), data scientists (DS), product managers (PM), operations staff (Ops), legal personnel (Le), designers (Des), and all participants (ALL), respectively, and each row represents one of the nine perspectives.}
\resizebox{\columnwidth}{!}{
\begin{tabular}{crrrrrrrr}
\hline
\textbf{Per.} & 
\textbf{\begin{tabular}[c]{@{}c@{}}Dev\end{tabular}} & 
\textbf{\begin{tabular}[c]{@{}c@{}}QA\end{tabular}} & 
\textbf{\begin{tabular}[c]{@{}c@{}}DS\end{tabular}} & 
\textbf{\begin{tabular}[c]{@{}c@{}}PM\end{tabular}} & 
\textbf{\begin{tabular}[c]{@{}c@{}}Ops\end{tabular}} & 
\textbf{\begin{tabular}[c]{@{}c@{}}Le\end{tabular}} & 
\textbf{\begin{tabular}[c]{@{}c@{}}Des\end{tabular}} & 
\textbf{\begin{tabular}[c]{@{}c@{}}ALL\end{tabular}} \\ \hline
Code          & 13.5\%       & 13.3\%      & 12.7\%      & 12.7\%      & 10.3\%       & 11.9\%      & 5.6\%        & 12.0\%       \\
Vis.          & 8.6\%        & 7.4\%       & 9.5\%       & 7.5\%       & 9.9\%        & 8.2\%       & 20.8\%       & 9.0\%        \\
Con.          & 10.6\%       & 13.1\%      & 12.4\%      & 9.2\%       & 13.0\%       & 13.3\%      & 6.9\%        & 12.1\%       \\
Per.          & 8.6\%        & 11.7\%      & 11.1\%      & 9.5\%       & 13.1\%       & 13.5\%      & 5.6\%        & 11.5\%       \\
Acc.          & 8.7\%        & 10.9\%      & 10.6\%      & 9.2\%       & 9.3\%        & 10.5\%      & 11.1\%       & 9.9\%        \\
Mai.          & 11.2\%       & 10.0\%      & 10.3\%      & 11.7\%      & 9.0\%        & 6.8\%       & 5.6\%        & 9.6\%        \\
FA            & 14.7\%       & 13.0\%      & 11.5\%      & 17.3\%      & 13.8\%       & 14.0\%      & 16.7\%       & 13.8\%       \\
VA            & 9.2\%        & 8.1\%       & 9.1\%       & 10.4\%      & 9.8\%        & 8.7\%       & 18.1\%       & 9.7\%        \\
CA            & 14.9\%       & 12.4\%      & 12.7\%      & 12.6\%      & 11.7\%       & 13.1\%      & 9.7\%        & 12.5\%       \\ \hline
\end{tabular}}
\label{User persona table}
\end{table}
}

\section{Conclusion}

In this paper, we have introduced WebCoderBench, a comprehensive benchmark consisting of 1,572 authentic user requirements and 24 evaluation metrics, providing an automated, comprehensive, and interpretable evaluation framework for the task of web app generation. Our evaluation results have revealed a narrowing effectiveness gap between open-source and closed-source LLMs, as well as the rapid evolution of the capabilities of LLMs, with no single model achieving dominant effectiveness across all the metrics. By incorporating the weights according to human preferences, WebCoderBench enables developers to optimize their models in a targeted manner based on interpretable evaluation results.


\section*{Acknowledgements}
We thank our industrial partner. We thank the anonymous reviewers for their constructive feedback. 

This work was partially supported by Fundamental and Interdisciplinary Disciplines Breakthrough Plan of the Ministry of Education of China (No. JYB2025XDXM118), National Natural Science Foundation of China (Grants No. U25A6023, 92464301), and the China National Petroleum Corporation Science and Technology Project under grant No. 2024ZZ46-06.

Tao Xie is also affiliated with the Key Laboratory of High Confidence Software Technologies (Peking University), Ministry of Education China; Institute of Systems for Advanced Computing at Fudan University, China; Shanghai Research Institute of Systems of Open Computing, China.

\section*{Limitations}

Our benchmark has six limitations. 

First, it currently evaluates only front‑end web applications and restricts implementations to native HTML. We consider that this design is reasonable because front‑end functionalities cover most real user needs, and the existing LLMs are still struggling to handle complex full‑stack development tasks. Native HTML also enables consistent source‑code analysis, whereas framework‑specific formats (e.g., React) complicate automated evaluation. Nonetheless, we plan to incorporate backend tasks and support common frameworks in future versions.

Second, our dataset and our implementation of the evaluation metrics cannot be released publicly due to internal legal constraints, because our dataset contains real user requirements. Closed-source data and evaluation also prevent data leakage and evaluation hacking. We plan to maintain and update our leaderboard actively. We also plan to prepare another batch of user requirements for open-sourcing and cross-validation, so as to verify the generalizability and robustness of our findings across different requirement sets.

Third, the dataset distribution can influence evaluation results. To reduce this limitation, we follow a standardized data‑collection pipeline to obtain sufficient and realistic user requirements.

Fourth, results can be affected by the reliability of the manual annotations. We mitigate this limitation by leveraging our industry partner’s mature crowd-sourcing workflow, where annotators have at least three years of development experience. We adopt a two‑stage labeling strategy in which LLMs generate labels and humans verify them. Critical steps, such as dataset filtering, are triple‑annotated, and majority voting is used to determine final labels.

Fifth, the design of evaluation metrics can impact the results. We aim to build a comprehensive, interpretable, and quantitative metric suite. Guided by industrial practices, public standards, and academic insights, we develop 24 metrics across 9 perspectives. We plan to further enrich the metric set as our future work.

Finally, metric correctness and stability can influence the evaluation. For rule‑based metrics, we manually inspect 50 scoring instances per metric and conduct additional code reviews to make sure that the implementations are as expected. For LLM‑as‑a‑judge metrics, we run each metric three times on 50 requirements and use the Mann–Whitney U test~\cite{mann1947test} to ensure that the score variance across runs is significantly lower than the variance across models. We follow the prior work~\cite{fullfront} for prompt design and use different LLMs for different metrics to mitigate preference leakage and bias~\cite{LLMjudgenotreliable,LLMjudgepreference,LLMjudgeuncertainty}.


\bibliography{custom}

@inproceedings{WebDevArena,
    title={{WebDev} {Arena}},
    url={https://web.lmarena.ai/},
	author={LMArena.ai},
	year={2025}
}

@inproceedings{DesignArena,
    title={Design {Arena}: {World's} largest crowdsourced benchmark for {AI-generated} design},
    url={https://www.designarena.ai/},
	author={Design Arena},
	year={2025}
}

@inproceedings{W3C,
    title={{W3C} standards and drafts},
    url={https://www.w3.org/TR/},
	author={World Wide Web Consortium},
	year={2025}
}

@inproceedings{Lighthouse,
    title={Introduction to {Lighthouse}},
    url={https://developer.chrome.com/docs/lighthouse/overview},
	author={Google},
	year={2025}
}

@inproceedings{HTMLHint,
    title={{HTML} {Hint}: {The} static code analysis tool you need for your {HTML}},
    url={https://htmlhint.com/},
	author={HTML Hint},
	year={2025}
}

@inproceedings{ESLint,
    title={{ESLint}: {Find} and fix problems in your {JavaScript} code},
    url={https://eslint.org/},
	author={ESLint},
	year={2025}
}

@inproceedings{GPT,
    title={Introducing {GPT‑5} for developers},
    url={https://openai.com/index/introducing-gpt-5-for-developers/},
	author={OpenAI},
	year={2025}
}

@inproceedings{DoubaoCoding,
    title={Doubao {Coding}},
    url={https://www.doubao.com/code/chat},
	author={ByteDance},
	year={2025}
}

@inproceedings{Doubao,
    title={Doubao},
    url={https://www.doubao.com/chat/},
	author={ByteDance},
	year={2025}
}

@inproceedings{Cursor,
    title={Cursor},
    url={https://cursor.com/cn},
	author={Anysphere},
	year={2025}
}

@inproceedings{Windsurf,
    title={Windsurf - the best {AI} for coding},
    url={https://windsurf.com/},
	author={Windsurf},
	year={2025}
}

@inproceedings{ClaudeCode,
    title={Claude {Code}},
    url={https://claude.com/product/claude-code},
	author={Anthropic},
	year={2025}
}

@inproceedings{TRAE,
    title={{TRAE} - collaborate with intelligence},
    url={https://www.trae.ai/},
	author={ByteDance},
	year={2025}
}

@inproceedings{Manus,
    title={Manus: {Hands} on {AI}},
    url={https://manus.im/},
	author={Manus},
	year={2025}
}

@inproceedings{Lovable,
    title={Lovable},
    url={https://lovable.dev/},
	author={Lovable},
	year={2025}
}

@inproceedings{Minimax-m2,
    title={{MiniMax} {M2} \& agent: {Ingenious} in simplicity},
    url={https://www.minimax.io/news/minimax-m2/},
	author={MiniMax},
	year={2025}
}

@inproceedings{Minimax-agent,
    title={{MiniMax} agent},
    url={https://agent.minimaxi.com/},
	author={MiniMax},
	year={2025}
}

@inproceedings{Claude,
    title={Overview | {Claude}},
    url={https://claude.com/product/overview},
	author={Anthropic},
	year={2025}
}

@inproceedings{StyleLint,
    title={{Stylelint}: {A} mighty {CSS} linter that helps you avoid errors and enforce conventions},
    url={https://stylelint.io/},
	author={Stylelint},
	year={2025}
}

@article{mann1947test,
  title={On a test of whether one of two random variables is stochastically larger than the other},
  author={Mann, Henry B and Whitney, Donald R},
  journal={The Annals of Mathematical Statistics},
  year={1947}
}

@article{surveysample,
  title={Sample size for survey research: {Review} and recommendations},
  author={Memon, Mumtaz Ali and Ting, Hiram and Cheah, Jun-Hwa and Thurasamy, Ramayah and Chuah, Francis and Cham, Tat Huei},
  journal={Journal of Applied Structural Equation Modeling},
  year={2020}
}

@article{borda,
  title={An axiomatization of {Borda's} rule},
  author={Young, H Peyton},
  journal={Journal of Economic Theory},
  year={1974}
}

@article{Qwen3,
  title={Qwen3 technical report},
  author={Yang, An and Li, Anfeng and Yang, Baosong and Zhang, Beichen and Hui, Binyuan and Zheng, Bo and Yu, Bowen and Gao, Chang and Huang, Chengen and Lv, Chenxu and others},
  journal={arXiv preprint arXiv:2505.09388},
  year={2025}
}

@article{starcoder,
  title={{StarCoder}: {May} the source be with you!},
  author={Li, Raymond and Allal, Loubna Ben and Zi, Yangtian and Muennighoff, Niklas and Kocetkov, Denis and Mou, Chenghao and Marone, Marc and Akiki, Christopher and Li, Jia and Chim, Jenny and others},
  journal={arXiv preprint arXiv:2305.06161},
  year={2023}
}

@article{Qwen2.5coder,
  title={{Qwen2.5-Coder} technical report},
  author={Hui, Binyuan and Yang, Jian and Cui, Zeyu and Yang, Jiaxi and Liu, Dayiheng and Zhang, Lei and Liu, Tianyu and Zhang, Jiajun and Yu, Bowen and Lu, Keming and others},
  journal={arXiv preprint arXiv:2409.12186},
  year={2024}
}

@article{deepseekcoder,
  title={{DeepSeek-Coder}: {When} the large language model meets programming -- The rise of code intelligence},
  author={Guo, Daya and Zhu, Qihao and Yang, Dejian and Xie, Zhenda and Dong, Kai and Zhang, Wentao and Chen, Guanting and Bi, Xiao and Wu, Yu and Li, YK and others},
  journal={arXiv preprint arXiv:2401.14196},
  year={2024}
}

@article{DeepSeek,
  title={{DeepSeek-V3} technical report},
  author={Liu, Aixin and Feng, Bei and Xue, Bing and Wang, Bingxuan and Wu, Bochao and Lu, Chengda and Zhao, Chenggang and Deng, Chengqi and Zhang, Chenyu and Ruan, Chong and others},
  journal={arXiv preprint arXiv:2412.19437},
  year={2024}
}

@article{Minimax-m1,
  title={{MiniMax-M1}: {Scaling} test-time compute efficiently with lightning attention},
  author={Chen, Aili and Li, Aonian and Gong, Bangwei and Jiang, Binyang and Fei, Bo and Yang, Bo and Shan, Boji and Yu, Changqing and Wang, Chao and Zhu, Cheng and others},
  journal={arXiv preprint arXiv:2506.13585},
  year={2025}
}

@inproceedings{MinHash,
  title={On the resemblance and containment of documents},
  author={Broder, Andrei Z},
  booktitle={Proceedings of the Compression and Complexity of Sequences},
  year={1997}
}

@article{MiniLM,
  title={{MiniLM}: {Deep} self-attention distillation for task-agnostic compression of pre-trained transformers},
  author={Wang, Wenhui and Wei, Furu and Dong, Li and Bao, Hangbo and Yang, Nan and Zhou, Ming},
  journal={Advances in Neural Information Processing Systems},
  year={2020}
}

@article{artifactsbench,
  title={{ArtifactsBench}: {Bridging} the visual-interactive gap in {LLM} code generation evaluation},
  author={Zhang, Chenchen and Li, Yuhang and Xu, Can and Liu, Jiaheng and Liu, Ao and Zhou, Changzhi and Deng, Ken and Wu, Dengpeng and Huang, Guanhua and Li, Kejiao and others},
  journal={arXiv preprint arXiv:2507.04952},
  year={2025}
}

@article{webuibench,
  title={{WebUIBench}: {A} comprehensive benchmark for evaluating multimodal large language models in {webUI-to-code}},
  author={Lin, Zhiyu and Zhou, Zhengda and Zhao, Zhiyuan and Wan, Tianrui and Ma, Yilun and Gao, Junyu and Li, Xuelong},
  journal={arXiv preprint arXiv:2506.07818},
  year={2025}
}

@article{webgenbench,
  title={{WebGen-Bench}: {Evaluating} {LLMs} on generating interactive and functional websites from scratch},
  author={Lu, Zimu and Yang, Yunqiao and Ren, Houxing and Hou, Haotian and Xiao, Han and Wang, Ke and Shi, Weikang and Zhou, Aojun and Zhan, Mingjie and Li, Hongsheng},
  journal={arXiv preprint arXiv:2505.03733},
  year={2025}
}

@article{fullfront,
  title={{FullFront}: {Benchmarking} {MLLMs} across the full front-end engineering workflow},
  author={Sun, Haoyu and Wang, Huichen Will and Gu, Jiawei and Li, Linjie and Cheng, Yu},
  journal={arXiv preprint arXiv:2505.17399},
  year={2025}
}

@article{frontendbench,
  title={{FrontendBench}: {A} benchmark for evaluating {LLMs} on front-end development via automatic evaluation},
  author={Zhu, Hongda and Zhang, Yiwen and Zhao, Bing and Ding, Jingzhe and Liu, Siyao and Liu, Tong and Wang, Dandan and Liu, Yanan and Li, Zhaojian},
  journal={arXiv preprint arXiv:2506.13832},
  year={2025}
}

@article{webbench,
  title={{Web-Bench}: {A} {LLM} code benchmark based on web standards and frameworks},
  author={Xu, Kai and Mao, YiWei and Guan, XinYi and Feng, ZiLong},
  journal={arXiv preprint arXiv:2505.07473},
  year={2025}
}

@article{designbench,
  title={{DesignBench}: {A} comprehensive benchmark for {MLLM-based} front-end code generation},
  author={Xiao, Jingyu and Wang, Ming and Lam, Man Ho and Wan, Yuxuan and Liu, Junliang and Huo, Yintong and Lyu, Michael R},
  journal={arXiv preprint arXiv:2506.06251},
  year={2025}
}

@article{zscore,
  title={Data mining: {A} preprocessing engine},
  author={Al Shalabi, Luai and Shaaban, Zyad and Kasasbeh, Basel},
  journal={Journal of Computer Science},
  year={2006}
}

@article{text2imagesurvey,
  title={A survey on quality metrics for text-to-image generation},
  author={Hartwig, Sebastian and Engel, Dominik and Sick, Leon and Kniesel, Hannah and Payer, Tristan and Poonam, Poonam and Glockler, Michael and Bauerle, Alex and Ropinski, Timo},
  journal={IEEE Transactions on Visualization and Computer Graphics},
  year={2025}
}

@article{Gemini,
  title={Gemini: {A} family of highly capable multimodal models},
  author={Team, Gemini and Anil, Rohan and Borgeaud, Sebastian and Alayrac, Jean-Baptiste and Yu, Jiahui and Soricut, Radu and Schalkwyk, Johan and Dai, Andrew M and Hauth, Anja and Millican, Katie and others},
  journal={arXiv preprint arXiv:2312.11805},
  year={2023}
}

@article{GLM,
  title={{GLM-4.5}: {Agentic}, reasoning, and coding (arc) foundation models},
  author={Zeng, Aohan and Lv, Xin and Zheng, Qinkai and Hou, Zhenyu and Chen, Bin and Xie, Chengxing and Wang, Cunxiang and Yin, Da and Zeng, Hao and Zhang, Jiajie and others},
  journal={arXiv preprint arXiv:2508.06471},
  year={2025}
}

@inproceedings{webcode2m,
  title={{WebCode2M}: {A} real-world dataset for code generation from webpage designs},
  author={Gui, Yi and Li, Zhen and Wan, Yao and Shi, Yemin and Zhang, Hongyu and Chen, Bohua and Su, Yi and Chen, Dongping and Wu, Siyuan and Zhou, Xing and others},
  booktitle={Proceedings of the ACM on Web Conference},
  year={2025}
}

@article{websight,
  title={Websight: {A} vision-first architecture for robust web agents},
  author={Bhathal, Tanvir and Gupta, Asanshay},
  journal={arXiv preprint arXiv:2508.16987},
  year={2025}
}

@article{design2code,
  title={Multimodal graph representation learning for website generation based on visual sketch},
  author={Vu, Tung D and Hoang, Chung and Hy, Truong-Son},
  journal={arXiv preprint arXiv:2504.18729},
  year={2025}
}

@article{vision2ui,
  title={{Vision2UI}: {A} real-world dataset with layout for code generation from {UI} designs},
  author={Gui, Yi and Li, Zhen and Wan, Yao and Shi, Yemin and Zhang, Hongyu and Su, Yi and Dong, Shaoling and Zhou, Xing and Jiang, Wenbin},
  journal={arXiv preprint arXiv:2404.06369},
  year={2024}
}

@article{sketch2code,
  title={{Sketch2Code}: {Evaluating} vision-language models for interactive web design prototyping},
  author={Li, Ryan and Zhang, Yanzhe and Yang, Diyi},
  journal={arXiv preprint arXiv:2410.16232},
  year={2024}
}

@article{web2code,
  title={{Web2Code}: {A} large-scale webpage-to-code dataset and evaluation framework for multimodal {LLMs}},
  author={Yun, Sukmin and Thushara, Rusiru and Bhat, Mohammad and Wang, Yongxin and Deng, Mingkai and Wang, Jinhong and Tao, Tianhua and Li, Junbo and Li, Haonan and Nakov, Preslav and others},
  journal={Advances in Neural Information Processing Systems},
  year={2024}
}

@inproceedings{pix2code,
  title={{Pix2Code}: {Generating} code from a graphical user interface screenshot},
  author={Beltramelli, Tony},
  booktitle={Proceedings of the ACM SIGCHI Symposium on Engineering Interactive Computing Systems},
  year={2018}
}

@article{LLMjudgenotreliable,
  title={Neither valid nor reliable? {Investigating} the use of {LLMs} as judges},
  author={Chehbouni, Khaoula and Haddou, Mohammed and Cheung, Jackie Chi Kit and Farnadi, Golnoosh},
  journal={arXiv preprint arXiv:2508.18076},
  year={2025}
}

@article{LLMjudgepreference,
  title={Preference leakage: {A} contamination problem in {LLM-as-a-judge}},
  author={Li, Dawei and Sun, Renliang and Huang, Yue and Zhong, Ming and Jiang, Bohan and Han, Jiawei and Zhang, Xiangliang and Wang, Wei and Liu, Huan},
  journal={arXiv preprint arXiv:2502.01534},
  year={2025}
}

@article{LLMjudgeuncertainty,
  title={Analyzing uncertainty of {LLM-as-a-judge}: {Interval} evaluations with conformal prediction},
  author={Sheng, Huanxin and Liu, Xinyi and He, Hangfeng and Zhao, Jieyu and Kang, Jian},
  journal={arXiv preprint arXiv:2509.18658},
  year={2025}
}

@inproceedings{arenaranking,
  title={Chatbot arena: {An} open platform for evaluating {LLMs} by human preference},
  author={Chiang, Wei-Lin and Zheng, Lianmin and Sheng, Ying and Angelopoulos, Anastasios Nikolas and Li, Tianle and Li, Dacheng and Zhu, Banghua and Zhang, Hao and Jordan, Michael and Gonzalez, Joseph E and others},
  booktitle={Proceedings of the 41st International Conference on Machine Learning},
  year={2024}
}

\clearpage

\setcounter{page}{1}
\appendix

\newtcolorbox{taskbox}[1][]{
  enhanced,
  breakable,
  colback=white,
  colframe=black,
  boxrule=0.8pt,
  arc=2pt,
  left=2pt,
  right=2pt,
  top=6pt,
  bottom=6pt,
  fonttitle=\bfseries\small,
  title=#1,
  before upper={\small},
}

\newtcolorbox{taskboxunbreak}[1][]{
  enhanced,
  colback=white,
  colframe=black,
  boxrule=0.8pt,
  arc=2pt,
  left=2pt,
  right=2pt,
  top=6pt,
  bottom=6pt,
  fonttitle=\bfseries\small,
  title=#1,
  before upper={\small},
}

\begin{center}
    {\large\bfseries Appendices}
\end{center}

\textbf{\large Table of Contents:}

\startcontents[appendix]
\printcontents[appendix]{l}{1}{}

\newpage

\section{Evaluation Metrics}
\label{Detailed Evaluation Metrics}

We present the detailed information of each evaluation metrics in Table~\ref{metrics table part1}, Table~\ref{metrics table part2}, and Table~\ref{metrics table part3}. In each of the tables, the ``Purpose'' column describes the rationale of each metric and the aspect that each metric aims to evaluate. The ``Implementation Detail'' column describes the implementation detail of each metric in natural language. The ``Score Calculation'' column describes how the final scores (ranging from 0 to 100) are computed, with scaling applied to ensure a uniform distribution and sufficient discriminability.

{\setlength{\tabcolsep}{2pt}
\begin{sidewaystable*}[htbp]
\caption{A detailed introduction to the purpose, implementation details, and score calculation formulas of our evaluation metrics (Part I).}
\centering
\small
\resizebox{\textwidth}{!}{
\begin{tabular}{lp{8cm}p{12cm}p{6cm}}
\hline
\textbf{Metric}                     & \textbf{Purpose}                                                                                                                                                                                                                                                                                                                                                                                                                                         & \textbf{Implementation   Detail}                                                                                                                                                                                                                                                                                                                                                                                                                                                                                                                                                                                                                                                                                                                                                                           & \textbf{Score   Calculation}                                                                                                                                                 \\ \hline
General   Functionality Correctness & This metric is used to evaluate the overall quality, correctness, and   compliance of JavaScript, CSS, and HTML code on a webpage.                                                                                                                                                                                                                                                                                                                       & This metric parses code files using a large language model and evaluates them   according to ten predefined code‑quality criteria, with each criterion scored   on a 0–10 scale.                                                                                                                                                                                                                                                                                                                                                                                                                                                                                                                                                                                                                           & score =   sum(scores\_of\_ten\_check\_rules)                                                                                                                                 \\  \hline
Best   Practices                    & This metric aims to assess the webpage’s adherence to web‑development   best practices, including but not limited to the use of HTTPS, avoiding   vulnerable JavaScript libraries, avoiding deprecated APIs, providing correct   document declarations, properly setting image aspect ratios, and making   reasonable permission requests.                                                                                                               & This metric integrates the Lighthouse auditing framework to construct a   best‑practices evaluation system consisting of nine core indicators. It   parses the Lighthouse‑generated JSON report and extracts the composite score   of the “best‑practices” category as the overall result.                                                                                                                                                                                                                                                                                                                                                                                                                                                                                                                 & score =   Lighthouse\_best\_practices\_score * 100                                                                                                                           \\  \hline
Error   Handling                    & This metric is designed to systematically evaluate the coverage of   exception‑handling mechanisms in JavaScript code by identifying functions   requiring error handling and comparing them with the actual implementation of   exception‑protection structures, thereby providing a quantitative basis for   code‑quality analysis.                                                                                                                    & This metric implements a multi‑level analysis pipeline to detect   exception‑handling mechanisms. It first extracts embedded JavaScript code   from the HTML document, including content inside script tags and   event‑handler code. It then identifies functions using regex patterns that   match multiple function‑definition forms (function declarations, function   expressions, arrow functions, etc.). A predefined heuristic rule set filters   functions requiring error handling based on name prefixes and risk‑operation   keywords (such as asynchronous operations or data‑storage APIs). For each   identified target function, syntax analysis is applied to detect whether   try‑catch blocks, Promise.catch, or similar exception‑handling structures are   present.                   & score = (1 - log2(1 + (no\_error\_handling\_num /   (require\_error\_handling\_num + 1)))) * 100                                                                        \\  \hline
Runtime Console Errors            & This metric evaluates HTML code quality by detecting error‑level messages   (such as ERROR, SEVERE) in the browser console to quantify code robustness.                                                                                                                                                                                                                                                                                                  & This metric uses Selenium WebDriver to construct a controlled browser‑testing   environment. The HTML is loaded in headless mode to simulate runtime   behavior. The core logic includes: counting lines of code via the file system   as a baseline parameter; capturing runtime console output via browser log   APIs; filtering logs using predefined error‑level categories (e.g., ERROR,   SEVERE). A state‑isolation mechanism clears cookies and storage before each   test to ensure independence.                                                                                                                                                                                                                                                                                                 & \begin{tabular}[c]{@{}l@{}}errors\_per\_1k =   (total\_errors / total\_lines) * 1000\\ raw\_score = 100 - (errors\_per\_1k * 20)\\ score = max(0, final\_score)\end{tabular} \\  \hline
Static   Syntax Checking (Linting)  & This metric aims to establish a multi‑language code‑quality evaluation   benchmark by performing static analysis to detect syntax and compliance   errors in HTML, CSS, and JavaScript files, thereby quantifying code quality   and providing standardized scoring.                                                                                                                                                                                     & This metric initializes a multilingual detection class by configuring supported   file‑extension sets and ignorable rule sets. Based on extension‑based   routing, it invokes htmlhint, eslint, and stylelint for syntax and compliance   checks. It parses each tool’s JSON output to extract error‑level   descriptions, rule identifiers, and location information, applies   rule‑filtering to exclude specified rules, and aggregates results into a   structured issue list and directory‑level statistics.                                                                                                                                                                                                                                                                                          & \begin{tabular}[c]{@{}l@{}}errors\_per\_1k =   (total\_errors / total\_lines) * 1000\\ raw\_score = 100 - (errors\_per\_1k * 20)\\ score=max(0, raw\_score)\end{tabular}     \\  \hline
General   Visual Experience         & This metric reflects the user's overall perception when using the target   web application, emphasizing evaluations of aesthetics, design quality, and   stylistic consistency between visual presentation and content.                                                                                                                                                                                                                                  & This metric uses a multimodal large model combined with specially designed prompts   to output n scores across n evaluation dimensions, each expressed as an x–y   interval.                                                                                                                                                                                                                                                                                                                                                                                                                                                                                                                                                                                                                               & score = min(sqrt(score\_generated\_by\_LLM)   * 10 + 20, 100)                                                                                                                \\  \hline
Component   Style Consistency       & This metric evaluates the use of card components in the target webpage,   focusing on whether card elements (such as titles, icons, and body text)   exhibit consistent and standardized structure and styling.                                                                                                                                                                                                                                          & This metric extracts all components in the webpage that contain “card”-related   identifiers and filters out nodes that fail structural requirements (such as   fewer than two children, inconsistent child‑node counts within a group, or   mismatched node hierarchy or tags). It then identifies parallel card   components and evaluates consistency and completeness of titles, icons, and   body‑text elements to detect inconsistent components.                                                                                                                                                                                                                                                                                                                                                    & score = (1 - log2(1 +   (inconsistent\_num / (total\_num + 1)))) * 100                                                                                                  \\ \hline
Icon   Style Consistency            & This metric evaluates whether icon assets used in the web application   belong to a consistent visual style, and whether inconsistencies exist in   size, line weight, or underlying shape.                                                                                                                                                                                                                                                              & This metric extracts all SVG elements in the HTML and groups them according to   shared container relationships. It evaluates icon‑library consistency across   six dimensions: icon‑set consistency, size uniformity, stroke‑width   uniformity, background‑shape consistency, background‑color consistency, and   background‑padding uniformity, and reports the number n of failed dimensions.                                                                                                                                                                                                                                                                                                                                                                                                          & score = max(0,100-25*failed\_dimention\_num)                                                                                                                                 \\

 \hline
\end{tabular}}
\label{metrics table part1}
\end{sidewaystable*}}

{\setlength{\tabcolsep}{2pt}
\begin{sidewaystable*}[htbp]
\caption{A detailed introduction to the purpose, implementation details, and score calculation formulas of our evaluation metrics (Part II).}
\centering
\small
\resizebox{\textwidth}{!}{
\begin{tabular}{lp{8cm}p{12cm}p{6cm}}
\hline
\textbf{Metric}                     & \textbf{Purpose}                                                                                                                                                                                                                                                                                                                                                                                                                                         & \textbf{Implementation   Detail}                                                                                                                                                                                                                                                                                                                                                                                                                                                                                                                                                                                                                                                                                                                                                                           & \textbf{Score   Calculation}                                                                                                                                                 \\ \hline
Layout   Consistency                & This metric assesses the visual alignment quality of webpage layouts by   detecting multi‑dimensional alignment of elements (e.g., edges, center lines)   and bottom alignment in multi‑column layouts, thereby quantifying layout   regularity.                                                                                                                                                                                                         & This metric performs alignment evaluation through the following steps: applying   Canny edge detection and morphological operations to extract element contours   from the webpage screenshot; filtering based on size, nesting, and   banner‑region masks to retain valid layout elements; quantifying positional   deviations of left/right edges, top/bottom boundaries, and center points; and   for multi‑column layouts, grouping elements into rows and comparing   upper/lower boundaries to check inter‑column alignment. The final score   combines multi‑dimensional alignment rates and bottom‑alignment issues,   accompanied by detailed diagnostics.                                                                                                                                        & score = (1 -   log2(1 + (total\_errors / (total\_elements + 1)))) * 100                                                                                                 \\  \hline
Layout   Sparsity                   & This metric evaluates spatial utilization and content density of webpage   screenshots by detecting the largest continuous homogeneous grayscale region   (blank rectangle) to quantify wasted whitespace. By calculating blank‑space   ratios and converting them into scores, it assesses whether the layout is   overly sparse or contains extensive ineffective whitespace, supporting   design‑quality evaluation and user‑experience optimization. & This metric identifies blank regions using a combination of grayscale‑tolerance   bucketing and maximal‑rectangle detection. The image is converted to   grayscale, and pixels are grouped into tolerance buckets (default threshold:   80), treating similar grayscale values as homogeneous and avoiding noise from   minor variations. For each bucket’s binary mask, a histogram‑based   maximal‑rectangle algorithm with a monotonic stack finds the largest   rectangular area. The largest rectangle across all buckets is recorded, and   its area ratio to the full image is computed as the blank‑space rate. A   nonlinear mapping converts the rate into a 0–100 score, penalizing layouts   with large blank areas. A visualization highlights the detected region with a   red bounding box. & score =   min(sqrt(100-sparsity\_rate)*10, 100)                                                                                                                              \\  \hline
Visual   Harmony Degree             & This metric evaluates the color harmony of webpage interfaces by   analyzing the spatial distribution and perceptual characteristics of dominant   colors in the image, providing objective metrics for assessing color   diversity, balance, and aesthetic coherence.                                                                                                                                                                                   & This metric evaluates color harmony using computer‑vision and color‑space   conversion techniques. It extracts dominant colors via K‑means clustering and   performs multi‑dimensional analysis in HSV space: color diversity via   Euclidean distance, saturation balance via mean and variance, brightness   contrast via value range and distribution concentration, hue harmony via   geometric relations on the hue circle (complementary, triadic, etc.), and   temperature balance via warm–cool ratios. A weighted blend (hue 0.30,   brightness 0.25, saturation 0.20, diversity 0.15, temperature 0.10) produces   the overall harmony score.                                                                                                                                                    & score = (diversity *   0.15 + saturation * 0.20 + brightness * 0.25 + hue * 0.30 + temperature *   0.10) * 100                                                               \\  \hline
Copywriting   Quality               & This metric systematically evaluates the overall quality of textual   content in HTML interfaces. Through multi‑dimensional quantitative analysis   (such as accuracy, clarity, conciseness, consistency, and user‑friendliness),   it ensures conformity with UX design norms and industry compliance standards.                                                                                                                                        & This metric extracts pure text from the HTML structure and evaluates it using   rules built from predefined domain‑specific vocabularies (such as system   operations, compliance terms, and technical terminology). Using a dimensional   scoring framework, it computes deviations from baseline across 15 quality   indicators (such as terminology consistency, sentence‑structure complexity,   and information hierarchy). Quantitative features include lexical matching,   statistical patterns (e.g., sentence length, punctuation mixing), and   structural conformance (e.g., heading hierarchy). Scores are aggregated into   an overall quality index.                                                                                                                                        & score =   Average(sub\_score\_1, sub\_score\_2, ..., sub\_score\_14)                                                                                                         \\  \hline
Media   Quality                     & This metric evaluates the quality of media resources (such as images and   videos) to ensure compliance with accessibility standards (e.g., WCAG) and   industry best practices, specifically by checking image clarity and video   playability.                                                                                                                                                                                                         & This metric parses the DOM and embedded scripts to extract all image and video   resource URLs. Image clarity is assessed via a Laplacian‑variance method;   video availability is assessed by verifying metadata readability and   first‑frame decodability.                                                                                                                                                                                                                                                                                                                                                                                                                                                                                                                                              & score = (clarity\_score   * 0.7) + (media\_accessbility\_score * 0.3)                                                                                                        \\  \hline
Placeholder   Quality               & This metric systematically evaluates the appropriateness of image   placement in webpage design by detecting four categories of   issues—placeholder image misuse, image distortion, repetitive use, and   improper SVG placeholders—and generating quantitative scores to measure   visual coherence.                                                                                                                                                   & This metric uses BeautifulSoup to extract image elements and contextual metadata.   It identifies placeholder images based on a predefined domain list, detects   distortion by comparing natural and rendered aspect ratios via remote size   probing and local attributes, identifies reuse by aggregating base URLs   within card containers, and assesses the appropriateness of SVG placeholders   in large containers using container‑size analysis and semantic context (with   special exemptions such as QR‑code cases).                                                                                                                                                                                                                                                                          & score = (1 - log2(1 +   (bad\_image\_num / (total\_image\_num + 1)))) * 100                                                                                                  \\

 \hline
\end{tabular}}
\label{metrics table part2}
\end{sidewaystable*}}

{\setlength{\tabcolsep}{2pt}
\begin{sidewaystable*}[htbp]
\caption{A detailed introduction to the purpose, implementation details, and score calculation formulas of our evaluation metrics (Part III).}
\centering
\small
\resizebox{\textwidth}{!}{
\begin{tabular}{lp{8cm}p{12cm}p{6cm}}
\hline
\textbf{Metric}                     & \textbf{Purpose}                                                                                                                                                                                                                                                                                                                                                                                                                                         & \textbf{Implementation   Detail}                                                                                                                                                                                                                                                                                                                                                                                                                                                                                                                                                                                                                                                                                                                                                                           & \textbf{Score   Calculation}                                                                                                                                                 \\ \hline
Resource   Validity                 & This metric evaluates the availability of embedded resources by checking   whether images, videos, audio files, and stylesheets referenced in HTML   documents are accessible, ensuring content completeness and user‑experience   quality. As part of the web‑quality benchmark, it quantifies resource load   success rates.                                                                                                                           & This metric parses the HTML DOM to extract two classes of resource paths: static   resources referenced via predefined tags (img, video, script, etc.) and URLs   embedded in JavaScript code (matched with regex patterns such as imageUrl and   src). It filters out template placeholders and special protocols (e.g.,   javascript:), then validates URLs by protocol: remote resources via layered   HTTP requests (HEAD first, fallback to GET, with redirect and SSL handling),   and local resources via file‑system checks. The system outputs structured   validation states and error attributions.                                                                                                                                                                                             & score = (1 - log2(1 +   (invalid\_resource\_num / (total\_resource\_num + 1)))) * 100                                                                                        \\  \hline
General   Performance               & This metric uses the Lighthouse tool to automate performance assessments   of HTML pages, quantifying performance metrics and generating detailed   performance reports.                                                                                                                                                                                                                                                                                 & This metric integrates Lighthouse’s performance‑analysis modules. It launches an   isolated local HTTP server to ensure a consistent environment, runs   Lighthouse with performance‑only collection and strict timeouts, and extracts   the Performance category score and five key metrics (FCP, LCP, TBT, CLS,   TTI), converting them into percentage‑scale values.                                                                                                                                                                                                                                                                                                                                                                                                                                    & score =   Lighthouse\_performance\_score * 100                                                                                                                               \\  \hline
Accessibility   Core Metrics        & This metric systematically evaluates accessibility compliance of webpage   content by detecting key accessibility features (such as alternative text,   keyboard navigation support, and semantic structure) for alignment with WCAG   standards, providing quantitative benchmarks for inclusive design.                                                                                                                                                & This metric invokes the Lighthouse accessibility module and parses the resulting   JSON report. It evaluates 11 core indicators, including alternative text,   button names, contrast ratios, heading structure, form labels, link   descriptions, ARIA compliance, document title, language declaration, and   viewport configuration.                                                                                                                                                                                                                                                                                                                                                                                                                                                                    & score =   Lighthouse\_accessibility\_score * 100                                                                                                                             \\  \hline

Cross-Browser   Compatibility       & This metric evaluates compatibility across different browser versions,   specifically examining the use of CSS properties and JavaScript APIs. It   identifies features that may fail in unsupported target browsers to quantify   compatibility levels.                                                                                                                                                                                                 & This metric integrates the MDN Browser Compatibility Database (BCD) as a reference   source to automate compatibility checks. It uses Playwright to load the   webpage and inject scripts to extract actual CSS properties (from stylesheets   and inline styles) and JavaScript APIs in use. These features are compared   against BCD data to determine support status for the target browser versions.                                                                                                                                                                                                                                                                                                                                                                                                  & score =   (compatible\_features / all\_features) * 100                                                                                                                       \\  \hline
Mobile   Device Compatibility       & This metric evaluates layout adaptability in mobile viewports by   detecting whether unexpected horizontal scrolling occurs at the document   root, thereby quantifying compliance with responsive‑design standards and   providing a baseline metric for layout viewport width.                                                                                                                                                                         & This metric hosts the HTML files on a local HTTP‑server cluster and uses   Playwright to emulate an iPhone 12 Pro environment. Injected JavaScript   computes the difference between scrollWidth and clientWidth of   documentElement to quantify horizontal overflow. Measurements are triggered   on DOMContentLoaded and stabilized with rendering‑frame synchronization and   microsecond‑level delays. Overflow pixels are mapped to an audit score.                                                                                                                                                                                                                                                                                                                                                  & score = max(0, 100 -   (horizontal\_overflow\_pixels) )                                                                                                                      \\  \hline
Code   Redundancy Rate              & This metric quantifies unused JavaScript and CSS resources to assess code   redundancy, providing measurable indicators for performance optimization and   code‑quality maintenance.                                                                                                                                                                                                                                                                     & This metric integrates relevant Lighthouse audits to extract redundancy metrics   for unused JavaScript and unused CSS rules. It parses the Lighthouse JSON   report to obtain numeric parameters such as potential byte savings and   resource details. If no redundant resources are detected, a full score is   returned; otherwise, the mean of relevant audit scores is used as the   composite rating.                                                                                                                                                                                                                                                                                                                                                                                               & score =   (Lighthouse\_unused\_javascript + Lighthouse\_unused\_css\_rules) / 2 * 100                                                                                        \\  \hline
Comment   Rate                      & This metric evaluates the readability and maintainability of HTML code by   analyzing comment coverage and converting the comment ratio into standardized   scoring.                                                                                                                                                                                                                                                                                     & This metric parses the HTML file to build a multilingual comment‑detection   framework. Using a line‑based statistical method and state‑machine algorithm,   it tracks multi‑line‑comment boundaries (HTML , CSS /* /, JavaScript // and /   */) and distinguishes comment lines from code lines. The comment‑line ratio   is used to compute a normalized evaluation score.                                                                                                                                                                                                                                                                                                                                                                                                                               & score =   min(sqrt(comment\_rate) * 10 + 60, 100)                                                                                                                            \\  \hline
Functional   Alignment              & This metric evaluates whether the HTML webpage satisfies user‑specified   functional modules and interactions.                                                                                                                                                                                                                                                                                                                                           & This metric uses a large language model to parse HTML code and validate it against   predefined functional criteria.                                                                                                                                                                                                                                                                                                                                                                                                                                                                                                                                                                                                                                                                                       & score =   (passed\_check\_point\_num / all\_check\_point\_num) * 100                                                                                                         \\  \hline
Visual   Alignment                  & This metric evaluates whether the HTML webpage meets user‑specified   visual‑design standards.                                                                                                                                                                                                                                                                                                                                                           & This metric uses a large language model to parse HTML code and validate it against   predefined visual inspection criteria.                                                                                                                                                                                                                                                                                                                                                                                                                                                                                                                                                                                                                                                                                & score = (passed\_check\_point\_num / all\_check\_point\_num) * 100                                                                                                           \\  \hline
Content   Alignment                 & This metric evaluates whether the HTML webpage satisfies user‑specified   content semantics and designated data.                                                                                                                                                                                                                                                                                                                                         & This metric uses a large language model to parse HTML code and validate it against   predefined content‑semantic criteria.                                                                                                                                                                                                                                                                                                                                                                                                                                                                                                                                                                                                                                                                                 & score = (passed\_check\_point\_num / all\_check\_point\_num) * 100                                                                                                           \\
 \hline
\end{tabular}}
\label{metrics table part3}
\end{sidewaystable*}}

\clearpage

\section{Unweighted Z-score Results}
\label{Detailed Evaluation Results}

We present the detailed and unweighted z-scores in Table~\ref{Detailed zscores}. 
Z-scores represent differences in standard deviations from the mean for each metric. We adopt z-scores to resize the raw scores of different metrics to a comparable scale, and then weight them according to our derived weights.

In this table, the values are greatly larger than those in Table~\ref{Main results table}, since these values are unweighted. The z-scores faithfully reflect the differences between models.

We can also rank the models according to their scores on each individual metric, yielding 24 sub‑leaderboards that reflect the models’ capabilities on specific metrics. Model developers can then target optimization efforts toward the metrics on which their models rank lower.

{\setlength{\tabcolsep}{2pt}
\begin{sidewaystable}[htbp]
\caption{Detailed unweighted z-scores for each metric and each model.}
\centering
\resizebox{\textwidth}{!}{
\begin{tabular}{clcccccccccccccccccccccccc}
\hline
\textbf{ID}          & \textbf{Model}           & \multicolumn{5}{c}{\textbf{\begin{tabular}[c]{@{}c@{}}Code\\ Quality\end{tabular}}} & \multicolumn{6}{c}{\textbf{\begin{tabular}[c]{@{}c@{}}Visual\\ Quality\end{tabular}}} & \multicolumn{4}{c}{\textbf{\begin{tabular}[c]{@{}c@{}}Content\\ Quality\end{tabular}}} & \textbf{\begin{tabular}[c]{@{}c@{}}Performance\\ Quality\end{tabular}} & \multicolumn{3}{c}{\textbf{Accessbility}}        & \multicolumn{2}{c}{\textbf{Maintainability}} & \textbf{\begin{tabular}[c]{@{}c@{}}Functional\\ Alignment\end{tabular}} & \textbf{\begin{tabular}[c]{@{}c@{}}Visual\\ Alignment\end{tabular}} & \textbf{\begin{tabular}[c]{@{}c@{}}Content\\ Alignment\end{tabular}} \\ \cline{3-26} 
\multicolumn{1}{l}{} &                          & 1          & 2          & 3          & 4          & \multicolumn{1}{c|}{5}          & 6         & 7        & 8        & 9        & 10       & \multicolumn{1}{c|}{11}       & 12             & 13             & 14             & \multicolumn{1}{c|}{15}             & \multicolumn{1}{c|}{16}                                                & 17      & 18      & \multicolumn{1}{c|}{19}      & 20         & \multicolumn{1}{c|}{21}         & \multicolumn{1}{c|}{22}                                                 & \multicolumn{1}{c|}{23}                                             & 24                                                                   \\ \hline
1                    & DeepSeek-R1-0528         & -0.3699    & 0.1256     & 0.1622     & 0.0938     & \multicolumn{1}{c|}{0.1041}     & -0.3956   & -0.0470  & -0.5862  & -0.0533  & 0.2614   & \multicolumn{1}{c|}{0.1952}   & -0.0512        & 0.1059         & 0.0128         & \multicolumn{1}{c|}{0.4508}         & \multicolumn{1}{c|}{0.0248}                                            & 0.1433  & 0.1725  & \multicolumn{1}{c|}{0.1076}  & 0.0422     & \multicolumn{1}{c|}{0.1701}     & \multicolumn{1}{c|}{-0.0697}                                            & \multicolumn{1}{c|}{0.0219}                                         & 0.0195                                                               \\
2                    & DeepSeek-V3.1            & -0.4223    & 0.1158     & 0.1518     & -0.0284    & \multicolumn{1}{c|}{0.0602}     & -0.8833   & 0.0801   & 0.2612   & 0.0505   & -0.0731  & \multicolumn{1}{c|}{0.0427}   & -0.0097        & -0.0654        & -0.1864        & \multicolumn{1}{c|}{-0.1057}        & \multicolumn{1}{c|}{0.0718}                                            & -0.0806 & -0.2108 & \multicolumn{1}{c|}{-0.1006} & 0.0593     & \multicolumn{1}{c|}{-0.0553}    & \multicolumn{1}{c|}{-0.0001}                                            & \multicolumn{1}{c|}{0.0368}                                         & -0.0252                                                              \\
3                    & DeepSeek-V3.1-Terminus   & -0.2846    & 0.1971     & 0.0791     & 0.1495     & \multicolumn{1}{c|}{0.1060}     & 0.7438    & 0.0655   & -0.2315  & 0.0350   & 0.2399   & \multicolumn{1}{c|}{0.1721}   & -0.0305        & -0.5671        & -0.1470        & \multicolumn{1}{c|}{0.1044}         & \multicolumn{1}{c|}{0.1203}                                            & 0.1017  & 0.1550  & \multicolumn{1}{c|}{0.0930}  & 0.1273     & \multicolumn{1}{c|}{0.3809}     & \multicolumn{1}{c|}{0.0369}                                             & \multicolumn{1}{c|}{-0.0029}                                        & 0.0024                                                               \\
4                    & DeepSeek-V3.1-Thinking   & -0.4038    & 0.1054     & 0.1582     & -0.0948    & \multicolumn{1}{c|}{0.0385}     & -0.8668   & 0.0488   & 0.2901   & 0.0781   & -0.0606  & \multicolumn{1}{c|}{0.0490}   & -0.0133        & 0.1260         & -0.1942        & \multicolumn{1}{c|}{-0.0728}        & \multicolumn{1}{c|}{0.1027}                                            & -0.0838 & -0.1078 & \multicolumn{1}{c|}{-0.1274} & 0.0760     & \multicolumn{1}{c|}{-0.0680}    & \multicolumn{1}{c|}{-0.0098}                                            & \multicolumn{1}{c|}{-0.0252}                                        & -0.0395                                                              \\
5                    & GLM-4.5                  & -0.2080    & 0.1642     & -0.0181    & 0.2018     & \multicolumn{1}{c|}{0.1080}     & 0.7800    & 0.0864   & 0.0392   & 0.0327   & 0.1503   & \multicolumn{1}{c|}{0.0780}   & -0.0276        & 0.8704         & 0.2424         & \multicolumn{1}{c|}{0.6038}         & \multicolumn{1}{c|}{0.0404}                                            & -0.0311 & 0.2128  & \multicolumn{1}{c|}{-0.0587} & 0.0725     & \multicolumn{1}{c|}{0.0558}     & \multicolumn{1}{c|}{0.1122}                                             & \multicolumn{1}{c|}{0.0885}                                         & 0.0361                                                               \\
6                    & Qwen3-Coder-Plus         & -0.3550    & -0.0543    & 0.1874     & 0.1608     & \multicolumn{1}{c|}{0.0849}     & -0.6834   & 0.1364   & -0.4843  & 0.0737   & 0.1757   & \multicolumn{1}{c|}{0.0451}   & 0.0008         & -0.3977        & -0.0472        & \multicolumn{1}{c|}{0.2005}         & \multicolumn{1}{c|}{0.1492}                                            & 0.1679  & -0.0163 & \multicolumn{1}{c|}{0.0883}  & 0.1934     & \multicolumn{1}{c|}{0.1257}     & \multicolumn{1}{c|}{-0.1105}                                            & \multicolumn{1}{c|}{-0.1105}                                        & -0.0861                                                              \\
7                    & Qwen3-235B-A22B-Instruct & -0.2647    & -0.1941    & 0.1561     & -0.0488    & \multicolumn{1}{c|}{-0.0281}    & 0.2797    & 0.1230   & -0.3037  & 0.1430   & -0.1480  & \multicolumn{1}{c|}{-0.1542}  & 0.0517         & -0.0613        & -0.0344        & \multicolumn{1}{c|}{-0.1425}        & \multicolumn{1}{c|}{0.0599}                                            & 0.0006  & -0.1347 & \multicolumn{1}{c|}{0.1377}  & 0.1295     & \multicolumn{1}{c|}{-0.0049}    & \multicolumn{1}{c|}{-0.0037}                                            & \multicolumn{1}{c|}{-0.0006}                                        & 0.0196                                                               \\
8                    & MiniMax-M2               & 0.5860     & 0.0799     & -0.4363    & 0.0143     & \multicolumn{1}{c|}{-0.2846}    & -0.3207   & -0.2503  & 0.0883   & 0.0294   & -0.0894  & \multicolumn{1}{c|}{0.1052}   & 0.0234         & -0.5651        & 0.1595         & \multicolumn{1}{c|}{0.1304}         & \multicolumn{1}{c|}{0.0355}                                            & -0.2779 & 0.1584  & \multicolumn{1}{c|}{-0.1204} & -0.0715    & \multicolumn{1}{c|}{0.4824}     & \multicolumn{1}{c|}{0.2272}                                             & \multicolumn{1}{c|}{0.1584}                                         & 0.1541                                                               \\ \hline
9                    & Gemini-2.5-Pro           & 0.3541     & -0.1512    & 0.0824     & -0.0155    & \multicolumn{1}{c|}{-0.0445}    & 0.4479    & -0.0888  & 0.5616   & -0.1118  & -0.2170  & \multicolumn{1}{c|}{-0.1729}  & 0.0612         & -0.2556        & 0.1294         & \multicolumn{1}{c|}{-0.0244}        & \multicolumn{1}{c|}{-0.1978}                                           & 0.0667  & 0.1441  & \multicolumn{1}{c|}{-0.0633} & -0.1404    & \multicolumn{1}{c|}{-0.2124}    & \multicolumn{1}{c|}{0.0645}                                             & \multicolumn{1}{c|}{0.1077}                                         & 0.0302                                                               \\
10                   & GPT-4o-2024-11-20        & -0.6459    & 0.0849     & 0.3867     & -0.3998    & \multicolumn{1}{c|}{0.1112}     & -0.2261   & 0.2475   & -0.4843  & 0.1350   & -0.6292  & \multicolumn{1}{c|}{-0.3277}  & 0.1031         & -0.2531        & 0.1347         & \multicolumn{1}{c|}{-0.5137}        & \multicolumn{1}{c|}{-0.2183}                                           & 0.2383  & -0.725  & \multicolumn{1}{c|}{0.1246}  & -0.0470    & \multicolumn{1}{c|}{-0.9680}    & \multicolumn{1}{c|}{-0.7851}                                            & \multicolumn{1}{c|}{-0.7003}                                        & -0.5588                                                              \\
11                   & GPT-5-Codex-High         & 0.6954     & -0.6779    & -0.1503    & 0.1408     & \multicolumn{1}{c|}{-0.0671}    & 1.0080    & 0.0008   & 0.1264   & -0.2603  & 0.2346   & \multicolumn{1}{c|}{-0.0971}  & -0.0988        & -0.2587        & 0.1573         & \multicolumn{1}{c|}{-0.1401}        & \multicolumn{1}{c|}{-0.1876}                                           & 0.2241  & 0.0950  & \multicolumn{1}{c|}{0.1973}  & -0.1243    & \multicolumn{1}{c|}{-1.3085}    & \multicolumn{1}{c|}{0.1495}                                             & \multicolumn{1}{c|}{0.1899}                                         & 0.1874                                                               \\
12                   & GPT-5-High               & 0.9259     & 0.1093     & -0.6730    & 0.0981     & \multicolumn{1}{c|}{-0.2977}    & 0.8959    & -0.2097  & -0.2034  & -0.2432  & 0.0554   & \multicolumn{1}{c|}{0.0790}   & -0.0662        & -0.5889        & 0.1692         & \multicolumn{1}{c|}{0.0545}         & \multicolumn{1}{c|}{0.1114}                                            & -0.4192 & 0.2234  & \multicolumn{1}{c|}{-0.2856} & -0.2415    & \multicolumn{1}{c|}{0.3605}     & \multicolumn{1}{c|}{0.3915}                                             & \multicolumn{1}{c|}{0.2497}                                         & 0.2745                                                               \\ \hline
13                   & Manus                    & 0.2278     & -0.0221    & -0.1865    & -0.2492    & \multicolumn{1}{c|}{0.2409}     & -0.4364   & 0.2511   & -0.1441  & 0.0068   & 0.0733   & \multicolumn{1}{c|}{-0.0144}  & -0.0222        & -0.0770        & 0.0752         & \multicolumn{1}{c|}{0.2568}         & \multicolumn{1}{c|}{-0.0490}                                           & -0.0310 & -0.4325 & \multicolumn{1}{c|}{0.0352}  & -0.2326    & \multicolumn{1}{c|}{0.4846}     & \multicolumn{1}{c|}{0.2342}                                             & \multicolumn{1}{c|}{0.0348}                                         & 0.0985                                                               \\
14                   & MiniMax Agent            & -0.0796    & 0.1917     & -0.2998    & -0.1372    & \multicolumn{1}{c|}{0.2409}     & -0.5171   & -0.4619  & -0.4263  & 0.0195   & -0.0248  & \multicolumn{1}{c|}{-0.0166}  & -0.1069        & 1.3456         & 0.2398         & \multicolumn{1}{c|}{0.6680}         & \multicolumn{1}{c|}{-0.1217}                                           & -0.0239 & -0.6377 & \multicolumn{1}{c|}{0.0761}  & -0.4752    & \multicolumn{1}{c|}{0.5712}     & \multicolumn{1}{c|}{0.1215}                                             & \multicolumn{1}{c|}{0.0812}                                         & 0.0020                                                               \\ \hline
\end{tabular}}
\label{Detailed zscores}
\end{sidewaystable}
}

\clearpage

\section{Analysis of Unscorable Cases}

In this section, we analyze the cases in which each evaluation metric fails to produce a valid score. When computing averaged scores and overall scores, we exclude unscorable cases rather than assigning them a score of zero, since extreme values would substantially distort the z‑score distribution. Given that the number of unscorable cases for each model does not differ greatly across metrics, we believe that computing the average z‑scores over only scorable cases is reasonable.

The number of unscorable cases for each evaluation metric and each model is shown in Table~\ref{Unscorable cases}. We conduct a manual analysis, and classify the unscorable cases into four categories.

First, no scorable content is found in the artifact (i.e., the denominator of the metric is zero). This category takes a major proportion among unscorable cases. Metrics such as Media Quality and Icon Style Consistency are computed using invalid media or inconsistent icons divided by the total number of media items or icons. However, many artifacts contain no media or icons at all, resulting in a zero denominator.

Second, external tool execution faces failures. Metrics such as Best Practices and Performance rely on external tools (e.g., Lighthouse) for scoring. For some artifacts, Lighthouse fails to produce a score due to issues such as timeouts and rendering errors.

Third, the LLM‑as‑a‑judge paradigm is disturbed by artifact content. In some cases, the LLM repeatedly generates the same character (e.g., ``>'') within the artifact, even repeating for hundreds of thousands of tokens. When scoring such artifacts, the large number of meaningless tokens prevents the LLM judge from reliably following the scoring prompt, resulting in failure.

Fourth, business logic causes the page to stay in a loading state or to exit immediately. In some cases, the artifact’s business logic keeps the page always loading, or terminates when the required local data files are missing. When these artifacts are evaluated using browser‑based automation, the process times out or fails to render, making them unscorable.

\begin{sidewaystable}[htbp]
\caption{Number of unscorable cases of each evaluation metric for each model.}
\centering
\resizebox{\textwidth}{!}{
\begin{tabular}{clcccccccccccccccccccccccc}
\hline
\textbf{ID}          & \textbf{Model}           & \multicolumn{5}{c}{\textbf{\begin{tabular}[c]{@{}c@{}}Code\\ Quality\end{tabular}}} & \multicolumn{6}{c}{\textbf{\begin{tabular}[c]{@{}c@{}}Visual\\ Quality\end{tabular}}} & \multicolumn{4}{c}{\textbf{\begin{tabular}[c]{@{}c@{}}Content\\ Quality\end{tabular}}} & \textbf{\begin{tabular}[c]{@{}c@{}}Performance\\ Quality\end{tabular}} & \multicolumn{3}{c}{\textbf{Accessbility}} & \multicolumn{2}{c}{\textbf{Maintainability}} & \textbf{\begin{tabular}[c]{@{}c@{}}Functional\\ Alignment\end{tabular}} & \textbf{\begin{tabular}[c]{@{}c@{}}Visual\\ Alignment\end{tabular}} & \textbf{\begin{tabular}[c]{@{}c@{}}Content\\ Alignment\end{tabular}} \\ \cline{3-26} 
\multicolumn{1}{l}{} &                          & 1          & 2          & 3           & 4          & \multicolumn{1}{c|}{5}         & 6       & 7          & 8         & 9         & 10      & \multicolumn{1}{c|}{11}      & 12            & 13              & 14             & \multicolumn{1}{c|}{15}             & \multicolumn{1}{c|}{16}                                                & 17    & 18    & \multicolumn{1}{c|}{19}   & 20         & \multicolumn{1}{c|}{21}         & \multicolumn{1}{c|}{22}                                                 & \multicolumn{1}{c|}{23}                                             & 24                                                                   \\ \hline
1                    & DeepSeek-R1-0528         & 0          & 15         & 99          & 8          & \multicolumn{1}{c|}{0}         & 3       & 1160       & 1455      & 646       & 3       & \multicolumn{1}{c|}{3}       & 2             & 1543            & 1406           & \multicolumn{1}{c|}{1268}           & \multicolumn{1}{c|}{15}                                                & 15    & 4     & \multicolumn{1}{c|}{4}    & 11         & \multicolumn{1}{c|}{0}          & \multicolumn{1}{c|}{0}                                                  & \multicolumn{1}{c|}{485}                                            & 91                                                                   \\
2                    & DeepSeek-V3.1            & 0          & 22         & 106         & 7          & \multicolumn{1}{c|}{0}         & 0       & 1307       & 1541      & 670       & 0       & \multicolumn{1}{c|}{0}       & 3             & 1506            & 1313           & \multicolumn{1}{c|}{1204}           & \multicolumn{1}{c|}{22}                                                & 22    & 2     & \multicolumn{1}{c|}{4}    & 20         & \multicolumn{1}{c|}{0}          & \multicolumn{1}{c|}{0}                                                  & \multicolumn{1}{c|}{485}                                            & 91                                                                   \\
3                    & DeepSeek-V3.1-Terminus   & 1          & 38         & 28          & 27         & \multicolumn{1}{c|}{0}         & 0       & 1263       & 1542      & 621       & 0       & \multicolumn{1}{c|}{0}       & 0             & 1549            & 1444           & \multicolumn{1}{c|}{1364}           & \multicolumn{1}{c|}{38}                                                & 38    & 27    & \multicolumn{1}{c|}{27}   & 11         & \multicolumn{1}{c|}{0}          & \multicolumn{1}{c|}{0}                                                  & \multicolumn{1}{c|}{485}                                            & 91                                                                   \\
4                    & DeepSeek-V3.1-Thinking   & 0          & 18         & 101         & 1          & \multicolumn{1}{c|}{0}         & 0       & 1309       & 1535      & 665       & 0       & \multicolumn{1}{c|}{0}       & 2             & 1501            & 1307           & \multicolumn{1}{c|}{1196}           & \multicolumn{1}{c|}{18}                                                & 18    & 1     & \multicolumn{1}{c|}{1}    & 17         & \multicolumn{1}{c|}{0}          & \multicolumn{1}{c|}{0}                                                  & \multicolumn{1}{c|}{485}                                            & 91                                                                   \\
5                    & GLM-4.5                  & 0          & 37         & 32          & 28         & \multicolumn{1}{c|}{0}         & 0       & 1211       & 1506      & 505       & 0       & \multicolumn{1}{c|}{0}       & 0             & 1425            & 1398           & \multicolumn{1}{c|}{1272}           & \multicolumn{1}{c|}{37}                                                & 37    & 27    & \multicolumn{1}{c|}{27}   & 10         & \multicolumn{1}{c|}{0}          & \multicolumn{1}{c|}{0}                                                  & \multicolumn{1}{c|}{485}                                            & 91                                                                   \\
6                    & Qwen3-Coder-Plus         & 0          & 13         & 122         & 3          & \multicolumn{1}{c|}{0}         & 0       & 1302       & 1558      & 665       & 0       & \multicolumn{1}{c|}{0}       & 0             & 1552            & 1441           & \multicolumn{1}{c|}{1408}           & \multicolumn{1}{c|}{13}                                                & 13    & 3     & \multicolumn{1}{c|}{3}    & 10         & \multicolumn{1}{c|}{0}          & \multicolumn{1}{c|}{0}                                                  & \multicolumn{1}{c|}{485}                                            & 91                                                                   \\
7                    & Qwen3-235B-A22B-Instruct & 7          & 45         & 161         & 33         & \multicolumn{1}{c|}{0}         & 0       & 1424       & 1558      & 874       & 0       & \multicolumn{1}{c|}{0}       & 0             & 1483            & 1320           & \multicolumn{1}{c|}{1175}           & \multicolumn{1}{c|}{45}                                                & 45    & 24    & \multicolumn{1}{c|}{24}   & 21         & \multicolumn{1}{c|}{0}          & \multicolumn{1}{c|}{1}                                                  & \multicolumn{1}{c|}{486}                                            & 97                                                                   \\
8                    & MiniMax-M2               & 5          & 18         & 61          & 3          & \multicolumn{1}{c|}{0}         & 0       & 1335       & 1441      & 735       & 0       & \multicolumn{1}{c|}{0}       & 2             & 1533            & 1429           & \multicolumn{1}{c|}{1287}           & \multicolumn{1}{c|}{18}                                                & 18    & 2     & \multicolumn{1}{c|}{2}    & 16         & \multicolumn{1}{c|}{0}          & \multicolumn{1}{c|}{2}                                                  & \multicolumn{1}{c|}{487}                                            & 94                                                                   \\ \hline
9                    & Gemini-2.5-Pro           & 2          & 54         & 175         & 21         & \multicolumn{1}{c|}{0}         & 0       & 1387       & 1460      & 865       & 0       & \multicolumn{1}{c|}{0}       & 1             & 1518            & 1345           & \multicolumn{1}{c|}{1299}           & \multicolumn{1}{c|}{54}                                                & 54    & 21    & \multicolumn{1}{c|}{21}   & 33         & \multicolumn{1}{c|}{0}          & \multicolumn{1}{c|}{3}                                                  & \multicolumn{1}{c|}{487}                                            & 92                                                                   \\
10                   & GPT-4o-2024-11-20        & 1          & 77         & 442         & 31         & \multicolumn{1}{c|}{0}         & 0       & 1516       & 1568      & 1236      & 0       & \multicolumn{1}{c|}{0}       & 0             & 1504            & 1368           & \multicolumn{1}{c|}{1232}           & \multicolumn{1}{c|}{77}                                                & 77    & 31    & \multicolumn{1}{c|}{31}   & 46         & \multicolumn{1}{c|}{0}          & \multicolumn{1}{c|}{0}                                                  & \multicolumn{1}{c|}{485}                                            & 91                                                                   \\
11                   & GPT-5-Codex-High         & 0          & 27         & 195         & 12         & \multicolumn{1}{c|}{0}         & 0       & 1130       & 1503      & 588       & 0       & \multicolumn{1}{c|}{0}       & 1             & 1533            & 1389           & \multicolumn{1}{c|}{1235}           & \multicolumn{1}{c|}{27}                                                & 27    & 12    & \multicolumn{1}{c|}{12}   & 15         & \multicolumn{1}{c|}{0}          & \multicolumn{1}{c|}{0}                                                  & \multicolumn{1}{c|}{485}                                            & 91                                                                   \\
12                   & GPT-5-High               & 0          & 17         & 32          & 9          & \multicolumn{1}{c|}{0}         & 1       & 1291       & 1377      & 556       & 0       & \multicolumn{1}{c|}{0}       & 1             & 1529            & 1411           & \multicolumn{1}{c|}{1382}           & \multicolumn{1}{c|}{17}                                                & 17    & 7     & \multicolumn{1}{c|}{7}    & 10         & \multicolumn{1}{c|}{0}          & \multicolumn{1}{c|}{0}                                                  & \multicolumn{1}{c|}{485}                                            & 91                                                                   \\ \hline
13                   & Manus                    & 1          & 10         & 2           & 3          & \multicolumn{1}{c|}{0}         & 0       & 120        & 144       & 60        & 0       & \multicolumn{1}{c|}{0}       & 1             & 145             & 137            & \multicolumn{1}{c|}{119}            & \multicolumn{1}{c|}{10}                                                & 10    & 0     & \multicolumn{1}{c|}{0}    & 10         & \multicolumn{1}{c|}{0}          & \multicolumn{1}{c|}{1}                                                  & \multicolumn{1}{c|}{44}                                             & 7                                                                    \\
14                   & MiniMax Agent            & 1          & 0          & 1           & 1          & \multicolumn{1}{c|}{0}         & 0       & 112        & 134       & 46        & 0       & \multicolumn{1}{c|}{0}       & 0             & 140             & 130            & \multicolumn{1}{c|}{95}             & \multicolumn{1}{c|}{0}                                                 & 0     & 0     & \multicolumn{1}{c|}{0}    & 0          & \multicolumn{1}{c|}{0}          & \multicolumn{1}{c|}{1}                                                  & \multicolumn{1}{c|}{44}                                             & 7                                                                    \\ \hline
\end{tabular}}
\label{Unscorable cases}
\end{sidewaystable}
\clearpage

\section{Questionnaire for Collecting Human Preferences}

The questionnaire that we use for collecting human-preference-aligned weights is presented as follows. We randomly reorder the options in each question to avoid bias.

\begin{taskbox}[Questionnaire Contents:]

We are studying the effectiveness of LLMs in the web application generation task. In this task, users provide the model with web‑development requirements (for example: “Help me build an XXX web application”), and the model generates the corresponding code of the web application.
\par\hspace{2em}\par
Please take the perspective of a real end user and, based on your personal preferences, rank the importance of the evaluation dimensions listed below. The ranking results will help us better understand users’ priorities in actual use.

\par\hspace{2em}\par
    1. When using a model to generate web applications, which perspectives do you care about the most? Please rank the following dimensions according to your level of importance.
    
        \hspace{2em}- Visual Quality: whether the webpage appears aesthetically pleasing and professional, and whether it adopts appropriate color combinations.
        
        \hspace{2em}- Code Quality: whether the webpage code is correctly implemented, free of bugs or errors, and compliant with engineering best practices.

        \hspace{2em}- Content Quality: whether the copywriting is clear to understand, the images and videos are high‑resolution, and the information is rich.
        
        \hspace{2em}- Performance Quality: whether the webpage loads smoothly and runs efficiently in practice.
        
        \hspace{2em}- Accessibility: whether the webpage benefits disabled people and displays correctly across different devices and browsers.
        
        \hspace{2em}- Maintainability: whether the webpage code is easy to read and modify.

        \hspace{2em}- Functional Alignment: whether the webpage functions and business logic are implemented according to your specifications.
                
        \hspace{2em}- Visual Alignment: whether the layout, colors, and appearance are implemented according to your specifications.
        
        \hspace{2em}- Content Alignment: whether the text, images, videos, and other resources in the webpage follow your specifications.

    \par\hspace{2em}\par
    
    2. When using a model to generate web applications, which aspects of visual aesthetics do you care about the most? Please rank the following dimensions by importance.

        \hspace{2em}- General Visual Experience: assessing the overall visual experience when using the webpage, emphasizing aesthetics, design quality, and consistency between style and content.

        \hspace{2em}- Component Style Consistency: assessing the usage of paratactic components and whether their internal elements (titles, icons, text, etc.) are stylistically consistent.

        \hspace{2em}- Icon Style Consistency: evaluating whether the icons used follow a unified style and whether inconsistencies exist in size, line weight, or background shape.
        
        \hspace{2em}- Layout Consistency: assessing how orderly the layout is, and whether components in rows or columns are properly aligned.
        
        \hspace{2em}- Layout Sparsity: evaluating the structural rationality of the layout and checking for excessively long or wide empty areas.
        
        \hspace{2em}- Visual Harmony Degree: evaluating the harmony of the webpage’s color scheme, including diversity, balance, and aesthetic coherence.

    \par\hspace{2em}\par
    3. When using a model to generate web applications, which aspects of code quality do you care about the most? Please rank the following dimensions by importance.

        \hspace{2em}- General Functionality Correctness: evaluating whether the implemented functions and business logic are correct.
        
        \hspace{2em}- Best Practices: evaluating adherence to web development best practices, including avoiding vulnerable JavaScript libraries, deprecated APIs, improper document declarations, or unreasonable permission requests.

        \hspace{2em}- Error Handling: evaluating the code’s ability to handle exceptional inputs (null values, errors, etc.).

        \hspace{2em}- Runtime Console Errors: assessing dynamic correctness by rendering the code in a browser and checking console logs for severe errors.
        
        \hspace{2em}- Static Syntax Checking: evaluating whether the code’s syntax and style are correct.

    \par\hspace{2em}\par
    4. When using a model to generate web applications, which aspects of content and media resources do you care about the most? Please rank the following dimensions by importance.

        \hspace{2em}- Copywriting Quality: assessing the overall quality of the text accuracy, clarity, brevity, contextual consistency, user‑friendliness, and compliance with UX and industry standards.
        
        \hspace{2em}- Media Quality: assessing the quality of media such as images and videos, including clarity and playability.
        
        \hspace{2em}- Placeholder Quality: evaluating whether placeholder images are used appropriately, without misuse, distortion, excessive repetition, or improper placement.

        \hspace{2em}- Resource Validity: assessing whether embedded resources (images, videos, audio, stylesheets) are accessible, previewable, and successfully loaded (e.g., no broken links or 404 errors).

    \par\hspace{2em}\par
    5. When using a model to generate web applications, which aspects of accessibility do you care about the most? Please rank the following dimensions by importance.

        \hspace{2em}- Accessibility Core Metrics: evaluating support for users with visual impairments and other groups, checking whether the readable description text is included, the keyboard navigation is supported, and the compliance with WCAG standards.
        
        \hspace{2em}- Cross‑Browser Compatibility: evaluating whether the webpage renders and functions correctly across different browsers.

        \hspace{2em}- Mobile Device Compatibility: assessing layout adaptability in mobile viewports and checking for overflow or abnormal rendering on mobile devices.
        
    \par\hspace{2em}\par
    6. When using a model to generate web applications, which aspects of code maintainability do you care about the most? Please rank the following dimensions by importance.

        \hspace{2em}- Code Redundancy Rate: assessing how much unnecessary or unused code is present, leading to tedious code.
        
        \hspace{2em}- Comment Rate: evaluating whether natural‑language comments are sufficient to aid developers in understanding the code.

\end{taskbox}

\clearpage

\section{Manual Labeling Guidelines}

This section presents our guidelines for human annotators to merge multi-turn requirements, filter the dataset using majority voting, build ground-truth checklists, and assign labels to each requirement.

\subsection{Merging Multi-Turn Requirements}

\begin{taskbox}[Guidelines for merging multi-turn requirements:]
    \textbf{Input}: Multi-turn user requirements, the merged requirements produced by an LLM.

    \textbf{Output}: The merged requirements that are checked and processed by human annotators.

    \textbf{Workflow}:
    \begin{itemize}
        \item Only turn-level additions or deletions are allowed; modification-only turns or turns without substantive requirement semantics should be removed.

        \item In cases of contradictory requirements, the later turn takes precedence; earlier conflicting turns should be discarded.
        
        \item Model-generated merged requirements are displayed in the ``summary'' column. Human annotators must validate outputs and correct issues according to Table~\ref{multi-round merging table}.
    \end{itemize}
\end{taskbox}

\begin{table*}[htbp]
\caption{Possible issues of requirements merged by the LLM.}
\resizebox{\textwidth}{!}{
\begin{tabular}{@{}lp{12cm}@{}}
\toprule
\textbf{Issue}                              & \textbf{Description}                                                                                                           \\ \midrule
Loss of Key Information            & Relevant content present in the original context is lost due to compression.                                                  \\
Redundant Content                  & After compression, repeated content still appears, or multi-turn requirements are merely concatenated without proper merging. \\
Hallucinated Content               & The model invents or incorrectly recalls information not present in the original dialogue.                                    \\
Content Rewriting /   Paraphrasing & User requirements are not fully preserved; the model rephrases or simplifies user expressions.                                  \\
Fail to Handle Conflicts           & Contradictory instructions across turns are not properly resolved.                                                              \\ \bottomrule
\end{tabular}}
\label{multi-round merging table}
\end{table*}

\subsection{Dataset Filtering}

\begin{taskbox}[Guidelines for dataset filtering:]
    \textbf{Input}: User requirements.

    \textbf{Output}: A True / False label for each requirement, indicating whether the requirement is usable.

    \textbf{Workflow}:
    Requirements that belong to the following cases should be marked as ``not usable''.
    \begin{enumerate}
    \item \textbf{Ambiguous or Logically Incoherent Requirements:} 
    The requirement is unclear or logically inconsistent. 
    Examples include:
    \begin{itemize}
        \item The user only uploads an HTML file without specifying any requirements.
        \item Requesting an ``infinite block map'' in HTML without further explanation.
    \end{itemize}

    \item \textbf{Missing Supplementary Data:} 
    The user's requirement lacks essential supplementary materials, such as images or links required to fulfill the request. 
    Example:
    \begin{itemize}
        \item ``Please generate a military training commemoration website with sections: 1. Title 2. Photo Wall (use only uploaded reference images, no external resources).'' 
        In this case, the required reference images are not provided.
    \end{itemize}

    \item \textbf{Non-native Web Scenarios or Non-Web Implementation Languages / Frameworks:} 
    The requirement specifies a context that is not native to standard web apps, or requires implementation in languages / frameworks outside of native HTML / CSS / JS. 
    Example:
    \begin{itemize}
        \item Using ESP32-S3 with a ST7789V display and FT6236U touch panel to create an interactive demo.
    \end{itemize}

    \item \textbf{Difficult-to-understand Requirements:} 
    The requirement cannot be quickly interpreted to identify the main functional requirements by an expert with front-end development experience. 
    Examples include:
    \begin{itemize}
        \item Not suitable: ``Toilet Man'' (ambiguous).  
        \item Suitable: ``Implement a Tetris game'' or ``Generate a Bomberman-style mini-game.''
    \end{itemize}
\end{enumerate}
\end{taskbox}

\subsection{Building Ground-Truth Checklists}

\begin{table*}[htbp]
\caption{Illustration of ground-truth checklists.}
\centering
\resizebox{\textwidth}{!}{
\begin{tabular}{p{4cm}p{4cm}p{10cm}}
\hline
\textbf{Dimension} & \textbf{Definition} & \textbf{Example} \\
\hline
Visual ground-truth checklist & Visual design requirements mentioned by users & ``I need a \textbf{red, industrial-style themed} webpage with a \textbf{blue} button below the form for submission.'' \\
\hline
Functional ground-truth checklist & 
Widget constraints mentioned by users

Functional points mentioned by users

Interaction actions mentioned by users
 & 
``I need \textbf{a table} displaying xxx information, with \textbf{a search box} above the table to input and filter results, and \textbf{a menu bar}.''

``The website can \textbf{search for types of marine sharks}.''

``I need the mouse to \textbf{hover over} the enter button for 3 seconds to enter the application; I need to \textbf{drag} the card to the schedule list to modify the schedule.''
\\
\hline
Content ground-truth checklist & Content to be displayed mentioned by users & ``I need to display \textbf{today's news} on the webpage.'' \\
\hline
\end{tabular}}
\label{GT table}
\end{table*}

\begin{taskbox}[Guidelines for building ground-truth checklists:]
    \textbf{Input}: User requirements, the functional, visual, and content ground-truth checklists generated by three distinct LLMs.

    \textbf{Output}: The functional, visual, and content ground-truth checklists validated and modified by human annotators.

    \textbf{Workflow}:
    You are required to validate and modify the ground-truth checklists generated by LLMs based on the user requirements. The possible operations include:
    \begin{itemize}
        \item Addition: Add requirements that are explicitly mentioned by the user, but missing in the LLM-generated checklists.

        \item Deletion: Remove requirements that are presented in the LLM-generated checklists but are not mentioned by the user and are unreasonable or should not appear in the final checklists.

        \item Retain: Retain requirements that at least two out of three models consider essential. Note: Different models can express the same requirement differently, and you should merge semantically equivalent items (e.g., ``bomb timing and explosion'' and ``bomb detonates on timer'').
    \end{itemize}

    The definitions and examples of each ground-truth checklist are shown in Table~\ref{GT table}.
\end{taskbox}

\subsection{Assigning Labels}
This section presents a detailed classification scheme and the underlying rationale for the requirements in the dataset.

\subsubsection{Input Modality}
The human annotators are asked to label each requirement according to Table~\ref{tab:input-modalities}.

\begin{table*}[htbp]
\caption{Classification of input modalities in user requirements.}
\centering
\renewcommand{\arraystretch}{1.25}
\setlength{\tabcolsep}{4pt}
\resizebox{\textwidth}{!}{
\begin{tabular}{p{0.20\textwidth} p{0.53\textwidth} p{0.35\textwidth}}
\hline
\textbf{Input Modality} & \textbf{Definition} & \textbf{Examples} \\
\hline
Text Only
& The user specifies requirements exclusively via textual description, without visual or structural references. This mode relies entirely on natural language to convey intended functionality, style, or content. 
& ``I want a black website header with three navigation links.'' \\
\hline
Text with Images 
& The user provides one or more images, design sketches, or screenshots, accompanied by a textual explanation specifying desired modifications or adaptations. The images can be used as either references or content. This approach leverages multimodal inputs to enhance specification clarity. 
& ``[Upload one or more website screenshots] Replicate the layout and color scheme of this page, but replace the text content with mine.'' \\
\hline
Text with URLs
& The user provides a specific web address (URL) as a reference point, requesting replication, adaptation, or stylistic emulation based on the referenced online resource. 
& ``Study the page style of \url{https://www.apple.com/mac/} and create a similar product introduction page for my offering.'' \\
\hline
\end{tabular}}
\label{tab:input-modalities}
\end{table*}

\subsubsection{Clarity of Requirement}
The human annotators are asked to label each requirement according to Table~\ref{tab:clarity-levels}. 

\begin{table*}[htbp]
\caption{Classification of requirement clarity levels.}
\centering
\renewcommand{\arraystretch}{1.25}
\setlength{\tabcolsep}{6pt}
\resizebox{\textwidth}{!}{
\begin{tabular}{p{0.04\textwidth} p{0.13\textwidth} p{0.28\textwidth} p{0.28\textwidth} p{0.25\textwidth}}
\hline
\textbf{Level} & \textbf{Designation} & \textbf{Description} & \textbf{Required Capability} & \textbf{Examples} \\
\hline
C1 & \textbf{Clear} 
& The user provides exceptionally concrete and detailed requirements, effectively resembling a concise specification document. This may include explicit functional elements, user interface components, and even prescribed interaction sequences. 
& \textit{Precise Execution}: The LLM is required to implement all specified details with exact fidelity, allowing minimal scope for autonomous interpretation. 
& ``Create a contact form with three required fields: `Name' (text input), `Email' (email input), and `Message' (text area). Below these fields, place a `Submit' button. Upon successful submission, clear the form and display: `Thank you for your message!'.'' \\
\hline
C2 & \textbf{Intermediate} 
& The user articulates a clear objective or core functionality while omitting most implementation details. The emphasis is on the desired outcome (\emph{what} is needed) rather than the process (\emph{how} to achieve it). 
& \textit{Interpretation and Completion}: The LLM must infer the essential goal and proactively complement missing specifications based on industry best practices or commonly observed design patterns (e.g., UI layout, interaction flows, error handling). 
& ``Develop a to-do list application.'' \newline
``I would like a weather forecast app.'' \\
\hline
C3 & \textbf{Vague} 
& The user presents only a loosely defined concept, intuition, or open-ended query, without a specific functional target. The requirement is exploratory in nature and encourages divergent thinking. 
& \textit{Creative Generation}: The LLM must engage in extensive association, reasoning, and ideation, and may proactively propose potential directions or features. In such cases, the notion of ``correctness'' is inherently non-deterministic. 
& ``Create a tool to improve my work efficiency.'' \newline
``Design an engaging homepage for my personal blog.'' \newline
``Suggest ways to make my photos look more stylish.'' \\
\hline
\end{tabular}}
\label{tab:clarity-levels}
\end{table*}

\subsubsection{Style of Expression}

The human annotators are asked to label each requirement according to Table~\ref{tab:expression-styles}. 

\begin{table*}[htbp]
\caption{Classification of expression styles.}
\centering
\renewcommand{\arraystretch}{1.25}
\setlength{\tabcolsep}{6pt}
\resizebox{\textwidth}{!}{
\begin{tabular}{p{0.04\textwidth} p{0.13\textwidth} p{0.24\textwidth} p{0.28\textwidth} p{0.29\textwidth}}
\hline
\textbf{Level} & \textbf{Designation} & \textbf{Description} & \textbf{Required Capability} & \textbf{Examples} \\
\hline
S1 & \textbf{Technical} 
& The user issues requirements in a precise, objective, and often technical manner, akin to a developer or product manager providing implementation directives. 
& \textit{Technical Terminology Comprehension}: The LLM must be capable of interpreting domain-specific jargon (e.g., ``SPA'', ``API'', ``responsive layout'', ``hook'') and accurately mapping such terminology to concrete implementation strategies. 
& ``Implement a SPA, containing a `Header` component and a reusable `Button` component.'' \newline
``Generate a RESTful API backend framework for the `user` entity, including CRUD endpoints.'' \\
\hline
S2 & \textbf{Colloquial} 
& The user communicates in everyday, informal language, similar to conversing with a friend. 
& \textit{Natural Language Understanding (NLU)}: The LLM must possess strong NLU capabilities to accurately extract core requirements and key entities from casual, idiomatic descriptions. 
& ``Hey, could you make me a small website to showcase photos of my cat so friends can view them?'' \newline
``I just want a simple expense tracker to record daily spending, with a monthly total view.'' \\
\hline
S3 & \textbf{Role-playing} 
& The user defines a context or adopts a role, thereby embedding the request within a rich narrative framework. This approach provides extensive situational information. 
& \textit{Contextual Comprehension and Empathy}: The LLM must adopt the specified role, understand the authentic pain points and latent needs apparent in the given scenario, and produce outputs aligned with the contextual demands. 
& ``As a fitness coach, I need an application to manage my clients' profiles and their training schedules.'' \newline
``Assume I am organizing a conference; I require a simple check-in page.'' \\
\hline
S4 & \textbf{Analogy} 
& The user describes requirements by drawing analogies to familiar applications or concepts. 
& \textit{Knowledge Transfer and Abstraction}: The LLM must identify the core functionalities and interaction patterns of the referenced analogy, abstract them, and adapt these elements to a new application domain. 
& ``Create a kanban board similar to Trello, but simpler.'' \newline
``I would like a photo filter feature similar to Instagram.'' \newline
``Develop a voting tool akin to WeChat group voting.'' \\
\hline
\end{tabular}}
\label{tab:expression-styles}
\end{table*}

\subsubsection{Artifact Complexity}
The human annotators are asked to label each requirement according to Table~\ref{tab:app-complexity}. 

\begin{table*}[htbp]
\caption{Classification of complexity levels of expected artifacts.}
\centering
\renewcommand{\arraystretch}{1.25}
\setlength{\tabcolsep}{4pt}
\resizebox{\textwidth}{!}{
\begin{tabular}{p{0.05\textwidth} p{0.11\textwidth} p{0.15\textwidth} p{0.19\textwidth} p{0.19\textwidth} p{0.19\textwidth} p{0.23\textwidth}}
\hline
\textbf{Level} & \textbf{Designation} & \textbf{Description} & \textbf{Function} & \textbf{Business Logic} & \textbf{UI/UX} & \textbf{Examples} \\
\hline
L1 & \textbf{Highly Simple} 
& A single, isolated functionality without data persistence, typically serving as a tool or static display. 
& Stateless single-function implementation, such as calculation, conversion, or plain text rendering. No data storage or backend involvement. 
& Direct linear logic: input → process → output, without conditional branching, multi-user roles, or state changes. 
& Minimalist interface: single page containing only essential I/O components (e.g., text field, button, label) and no navigation. 
& ``Generate a Celsius-to-Fahrenheit temperature converter.'' \newline
``Create a page displaying `Hello, World!'.'' \\
\hline
L2 & \textbf{Simple} 
& A basic CRUD application centered on a single core entity. 
& Single-entity CRUD: manages one main object (e.g., to-do item, note) with fundamental data persistence (local storage or simple database). 
& Simple state management: create, read, update, and delete operations for a single entity without complex relations or access control mechanisms. 
& Single-page dynamic application (SPA): operations are handled entirely within one page by component showing/hiding/updating, including lists and simple forms. 
& ``Build a to-do list application that allows adding, deleting, and marking tasks as completed.'' \newline
``Create a simple note-taking application with list and view functionality.'' \\
\hline
L3 & \textbf{Medium} 
& Involves multiple interrelated functional modules, or includes simple workflows and role distinctions. 
& Multi-module / multi-entity relationships: at least two associated entities (e.g., users and articles, products and categories). May involve basic third-party API calls (e.g., weather data). 
& Conditional and role-based logic: distinguishes between simple user roles (e.g., administrator vs. standard user). Supports basic workflows (e.g., article publication requiring review), with well-defined data validation rules. 
& Multi-page / multi-view navigation: incorporates multiple pages or views (e.g., homepage, detail page, admin panel) with structured routing and navigation (menus, tabs). 
& ``Develop a simple blogging system with user registration/login, allowing users to publish articles and administrators to review them.'' \newline
``Create a book management system that enables adding book information (title, author) and viewing by author categories.'' \\
\hline
L4 & \textbf{Complex} 
& Comprises complex business processes, multi-system integration, and rich interactive interfaces. 
& Integration with multiple systems/services closely tied to core business (e.g., payment gateways, mapping services, SMS verification). Requires handling of asynchronous operations and data aggregation. 
& Complex multi-step workflows: business logic involves sequential processes and state changes (e.g., e-commerce order: cart → address → payment → confirmation). Includes advanced permission handling and validation rules. 
& Dynamic, feature-rich interfaces: advanced forms, data filtering, sorting, visualizations. Real-time or near-real-time updates based on user actions or backend data. Requires responsive design. 
& ``Develop an online food ordering application where users browse menus, add items to a cart, pay via Alipay, and view order status.'' \newline
``Build a project task board enabling creation of task cards and drag-and-drop movement between `To Do', `In Progress', and `Done' lists.'' \\
\hline
L5 & \textbf{Highly Complex} 
& Enterprise-level or platform-scale application requiring high concurrency, real-time collaboration, and advanced algorithmic or data processing capabilities. 
& Platform/system-level functionality: scalable architecture supporting multi-tenancy, real-time communication (WebSocket), or complex background operations (e.g., data analytics, ML model invocation). 
& Highly complex business rules and finite-state machines: fine-grained access control, risk management, financial computation, or multi-party synchronization logic. 
& Highly dynamic and collaborative UI/UX: supports real-time multi-user collaborative actions (e.g., collaborative document editing, design tools). Includes advanced data visualization and customizable layouts, requiring high performance and UX quality. 
& ``Build a Trello-like team collaboration platform with boards, lists, and cards, supporting drag-and-drop and real-time synchronization of team member actions.'' \newline
``Create a basic online code editor with syntax highlighting and real-time collaborative editing among multiple users in the same session.'' \\
\hline
\end{tabular}}
\label{tab:app-complexity}
\end{table*}

\clearpage

\section{Prompts for LLMs}
This section presents the prompts that we use to leverage LLMs in our dataset construction process, data analysis, generating artifacts, and conducting evaluations.

\subsection{Prompts for Merging Multi-Turn Requirements}

During the dataset construction process, we first merge multi-turn requirements to single-turn ones using the LLM, and then let human annotators conduct validation and modification. The prompts that we use are as follows.

\begin{taskbox}[Prompts for merging multi-turn requirements:]
You are a Requirement Analyst. Your task is to process a multi-turn conversation record regarding ``application generation'' requirements, and merge the contents of multiple turns.

**Task Objective**:

1. Determine — except for the first turn — whether each subsequent turn is: Functionality Addition (e.g., ``Add XX functionality''), Functionality Fix (e.g., ``Fix XX issue''), Non-functional Description (e.g., ``Confirm requirement'', ``Start generation'', ``Continue''). Keep only the turns of the Functionality Addition type, and exclude all others.

2. Merge the first turn with the subsequent turns that are Functionality Addition, keeping the original description intact as much as possible, including any typos, without altering the original content — only performing a simple merge.

3. Between merged sentences, you may add or slightly modify a few words or sentences to make the text coherent and free of obvious merge traces.

4. If the original content contains JSON, only modify the value of the ``text'' field.

Only output the merged content. Do not provide any explanations or additional text.

Now, please merge based on the following multi-turn conversation content:

\{User Requirements\}
\end{taskbox}

\subsection{Prompts for Generating Classification Labels}

To analyze the statistics of our dataset, we first employ LLMs to generate detailed labels for each requirement, and then ask human annotators to check and revise these labels. The prompts that we use are shown in Figure~\ref{Fig:LabelPE}.

\begin{figure*}[htbp]
    \begin{taskboxunbreak}[Prompts for generating classification labels:]
**Role**

\hspace{2em}You are an experienced AI evaluation specialist, possessing an integrated perspective that combines the expertise of a software architect, senior product manager, and user researcher. You excel at accurately and objectively analyzing user requirements, and classifying them according to a rigorous set of standardized rules.

**Task**

\hspace{2em}Your task is to examine the content of [USER\_PROMPT\_TO\_ANALYZE] and evaluate it strictly according to the three dimensions defined in the ``Dimensions \& Rubrics'' section: Artifact Complexity, Prompt, and Artifact Type.
For each dimension, you must provide a concise rationale for the assigned label and produce a complete analysis output in the specified JSON format.

**Dimensions \& Rubrics**

**Dimension 1: Artifact Complexity**

**Rules:**

\hspace{2em}1. Assess the following three sub-dimensions separately: Functional Complexity, Business Logic, User Interaction. Assign a level from L1 to L5 for each.

\hspace{2em}2. Determine the final overall complexity level according to the ``Highest Level Principle'' — the sub-dimension with the highest level determines the final rating.

\resizebox{\textwidth}{!}{\begin{tabular}{|p{0.05\textwidth}|p{0.3\textwidth}| p{0.3\textwidth}| p{0.3\textwidth}|}
\hline
\textbf{Level} & \textbf{Functional Complexity}                                               & \textbf{Business Logic}                                                & \textbf{User Interaction}                               \\ \hline
L1             & Single, stateless functionality; no data storage.                            & Linear direct logic (input → output).                                  & Minimal interface (single page; basic components).      \\ \hline
L2             & CRUD operations on a single entity; basic data persistence.                  & Simple state management for a single entity.                           & Single-page dynamic app (lists, forms).                 \\ \hline
L3             & Multiple modules/entities; simple third-party API integration.               & Conditional and role-based logic (e.g., admin vs user).                & Multi-page/view navigation with routing.                \\ \hline
L4             & Deep integration with multiple systems/services (e.g., payments, maps).      & Complex multi-step workflows (e.g., e-commerce ordering).              & Rich dynamic interactions (filtering, sorting, charts). \\ \hline
L5             & Platform-level functionality (e.g., real-time communication, multi-tenancy). & Highly complex business rules and state machines (e.g., risk control). & Highly dynamic and collaborative UI (e.g., co-editing). \\ \hline
\end{tabular}}

**Dimension 2: Prompt Style**

**Rules:**

\hspace{2em}1. Clarity: Choose from {C1, C2, C3}.

\hspace{2em}2. Expression Style: Choose from {S1, S2, S3, S4}.

\hspace{2em}3. The final result must include both labels.

**2.1 Clarity**

\hspace{2em}C1: Clear \& Specific — Explicit, detailed requirements akin to a small specifications document.

\hspace{2em}C2: Goal-Oriented — Defines clear objectives but omits implementation details.

\hspace{2em}C3: Vague \& Exploratory — Expresses a broad idea or open-ended question only.

**2.2 Expression Style**

\hspace{2em}S1: Instructional/Technical — Precise, objective, and possibly technical language.

\hspace{2em}S2: Colloquial/Natural Language — Everyday, informal wording.

\hspace{2em}S3: Scenario/Role-playing — Describes requirements via set scenarios or role assumptions.

\hspace{2em}S4: Analogy/Heuristic — Expresses ideas through analogy with well-known applications or concepts.

**Dimension 3: Application Type**

\hspace{2em}Rules: Select the most appropriate type from the following list:

\hspace{2em}E-commerce, Online Education Platform, Healthcare Platform, Travel Services, Financial Services, News Media Platform, Entertainment/Gaming, Multimedia Platform, Corporate Website, Online Office Platform, Enterprise Back-office Management, AI Application, Smart Device Interaction, Social Media Platform, Forum Website, Personal Website, Public Service Platform, Utility Website, Data Visualization, Science Popularization Demonstration

**Output Format**

\hspace{2em}Your output must be a JSON object with the following structure. No explanations or text should appear outside the JSON block.

\{\{
  "application\_complexity": \{\{
    "final\_level": "L\_x\_",
    "final\_justification": "Core rationale for determining final level.",
    "dimensional\_analysis": [
      \{\{
        "dimension": "Functional Complexity",
        "level": "L\_x\_",
        "justification": "Concise rationale for this dimension."
      \}\},
      \{\{
        "dimension": "Business Logic",
        "level": "L\_x\_",
        "justification": "Concise rationale for this dimension."
      \}\},
      \{\{
        "dimension": "User Interaction",
        "level": "L\_x\_",
        "justification": "Concise rationale for this dimension."
      \}\}
    ]
  \}\},
  "prompt\_style": \{\{
    "clarity": \{\{
      "level": "C\_x\_",
      "justification": "Concise rationale for clarity rating."
    \}\},
    "expression": \{\{
      "level": "S\_x\_",
      "justification": "Concise rationale for expression style rating."
    \}\}
  \}\},
  "application\_type": \{\{
    "type": "Selected type from list",
    "justification": "Concise rationale for type selection."
  \}\}
\}\}

[USER\_PROMPT\_TO\_ANALYZE]

---

\{single-turn requirements\}

---

    \end{taskboxunbreak}
\caption{Prompts for generating classification labels.}
\label{Fig:LabelPE}
\end{figure*}

\subsection{Prompts for Generating Ground-Truth Checklists}

In order to check the alignment of the generated web apps with the corresponding requirements, we let three LLMs generate ground-truth checklists separately, and then ask human annotators to decide the final checklists for each requirement in our dataset. The prompts that we use are shown in Figure~\ref{Fig:GTPE} and Figure~\ref{Fig:GTPECont}.

\begin{figure*}[htbp]
    \begin{taskboxunbreak}[Prompts for generating ground-truth checklists:]
\# **[System Role]**  
You are a senior requirements analyst and evaluation standards expert, responsible for understanding user needs. We currently have a real-world user-provided requirement for a **web frontend application**. This requirement may be vague or detailed.  

In order to design and deliver a **fully functional, visually appealing, and content-rich product** that satisfies the user, we must break down the requirement into three structured dimensions: **Functional**, **Visual**, and **Content**.  

You must generate **Ground-Truth requirement points** in these three dimensions that correspond to the user's stated needs.

Structured dimensions are as follows:  

\hspace{2em}- **Functional**  

\hspace{2em}- **Visual**  

\hspace{2em}- **Content**  

Please strictly follow the rules below and **base your analysis solely on the user requirement text**. Do not make subjective assumptions or expand beyond what is explicitly stated.

---

\#\# **[Task Objective]**  
Output the following types of **GroundTruth requirement points**:

\#\#\# 1. **Functional**  

\hspace{2em}The operations, workflows, or system functions mentioned or implied in the requirement (*“What should the system do and how should it interact with the user?”*).

\#\#\# 2. **Visual**  

\hspace{2em}Experience-related requirements concerning theme colors, responsive layout, animation effects (*“What should the system look like, and what mandatory components must be included?”*).

\#\#\# 3. **Content**  

\hspace{2em}Page language type, videos, images, music, text copy, data sources, and other materials related to display or experience (*“What content should the system present?”*).

---

\#\# **[Decomposition Rules]**

\#\#\# **Functional Dimension** (Functional)

**Goal:** Extract the **Minimal Functional Set (MFS)** required to fulfill the user’s need.  

**Criteria:**  

\hspace{2em}1. **Explicit mention first**: If requirement includes operational verbs (e.g., *upload, play, share*), directly split into requirement points.  

\hspace{2em}2. **Implicit completion**: If requirement is an abstract objective (*e.g., “create a file sharing platform”*), extract the minimal functional set to achieve it: 

\hspace{2em}\hspace{2em}- Upload files  
   
\hspace{2em}\hspace{2em}- Generate sharing link  
   
\hspace{2em}\hspace{2em}- Access link to download  
   
\hspace{2em}3. **No divergence**: Do not infer features not mentioned (*e.g., “points system”, “admin dashboard”*). 

\hspace{2em}4. **Consistent granularity**: Requirement point should be independently developable and testable (includes input, processing, output).  

\hspace{2em}5. Do not include programming language or framework requirements (*e.g., “must use Vue framework”*).  

**Example:**  

> “The system should be able to display product videos online.” → [Upload video], [Play video]  

> “Users can upload and share files.” → [Upload file], [Generate sharing link], [Download via link]  

---

\#\#\# **Visual Dimension** (Visual)

**Goal:** Identify specific user demands for visual experience.  

**Criteria:**  

\hspace{2em}1. **Explicit mention first**: If requirement mentions colors or theme description (*e.g., “blue and white”, “industrial style”*), extract as requirement point.  

\hspace{2em}2. Focus on **theme colors, responsive layout, animation effects**.  

\hspace{2em}3. Do not mention basic UI elements (buttons, input boxes, tables, etc.) unless explicitly stated. 

\hspace{2em}4. If mentions *style*, *brand colors*, *animation effects*, *adaptation for mobile/PC*, it is considered a visual element.  

**Example:**  

> “Overall theme should be blue and white” → [Theme color: blue \& white]  

> “Interface must adapt for both mobile and desktop” → [Responsive layout]  

> “Page transitions must have fade-in/out effects” → [Animation: fade-in/out]  

---
    \end{taskboxunbreak}
\caption{Prompts for generating ground-truth checklists.}
\label{Fig:GTPE}
\end{figure*}

\begin{figure*}[htbp]
    \begin{taskboxunbreak}[Prompts for generating ground-truth checklists cont.:]
\#\#\# **Content Dimension** (Content)

**Goal:** Extract requirement points about content to be displayed.  

**Criteria:**  

\hspace{2em}1. **Explicit mention first**: If requirement lists specific media or content (*e.g., “today’s news”, “images”*), directly extract. If content (images, text, etc.) is provided by user, emphasize “provided by user”.  

\hspace{2em}2. Extract page language type, images, videos, audio, music, text copy, data sources, etc. 

\hspace{2em}3. Exclude logical text (*e.g., prompts, error messages, guiding instructions*).  

**Example:**  

> “Display company promotional video and background music.”  → [Video: Company promo], [Music: Background track]  

---

\#\# **[Output Format]**  

The output must be in JSON format (not Markdown JSON) with the following structure:

\{

  \hspace{2em}"functionals": [
  
    \hspace{2em}\hspace{2em}\{
    
      \hspace{2em}\hspace{2em}"type": "functional",
      
      \hspace{2em}\hspace{2em}"name": "Functional requirement point name",
      
      \hspace{2em}\hspace{2em}"description": "Brief description of the function and its application scenario"
      
    \hspace{2em}\hspace{2em}\}
    
  \hspace{2em}],
  
  \hspace{2em}"visuals": [
  
    \hspace{2em}\hspace{2em}\{
    
      \hspace{2em}\hspace{2em}"type": "visual",
      
      \hspace{2em}\hspace{2em}"name": "Visual requirement point name",
      
      \hspace{2em}\hspace{2em}"description": "Describe the purpose and presentation of the visual element (only color/responsive/animation)"
      
    \hspace{2em}\hspace{2em}\}
    
  \hspace{2em}],
  
  \hspace{2em}"contents": [
  
    \hspace{2em}\hspace{2em}\{
    
      \hspace{2em}\hspace{2em}"type": "content",
      
      \hspace{2em}\hspace{2em}"name": "Content requirement point name",
      
      \hspace{2em}\hspace{2em}"description": "Describe the purpose and details of the content (video/image/music/copy/data source, etc.)"
      
    \hspace{2em}\hspace{2em}\}
    
  \hspace{2em}],
  
  \hspace{2em}"summary": \{
  
    \hspace{2em}\hspace{2em}"functional": [
    
      \hspace{2em}\hspace{2em}"Function A",
      
      \hspace{2em}\hspace{2em}"Function B"
      
    \hspace{2em}\hspace{2em}],
    
    \hspace{2em}\hspace{2em}"visual": [],
    
    \hspace{2em}\hspace{2em}"content": []
    
  \hspace{2em}\}
  
\}

    \end{taskboxunbreak}
\caption{Prompts for generating ground-truth checklists continue.}
\label{Fig:GTPECont}
\end{figure*}

\subsection{Prompts for Generating Artifacts}

\begin{taskbox}[Prompts for generating artifacts using LLMs:]
You are a professional web front‑end application engineer and designer. You will receive user requirements for front‑end web pages and write web page code to fulfill those requirements.

Note:

\hspace{2em}1. Your output should only include the code itself, with no additional explanations.

\hspace{2em}2. You may only use native front‑end languages (HTML, JS, CSS) to build the page.
\end{taskbox}

\begin{taskbox}[Prompts for generating artifacts using LLM-based agents:]
You are a professional web front‑end application engineer and designer. You will receive user requirements for front‑end web pages and write web page code to fulfill those requirements.

\#\#Delivery Requirements

\hspace{2em}1. You must implement the requirements using only native front‑end languages (HTML, JS, CSS).

\hspace{2em}2. If the implementation can be done in a single file, then you may deliver only one HTML file.

\hspace{2em}3. If the implementation requires multiple files, then you may deliver only three types of files: HTML files, CSS files, and JS files.

\hspace{2em}4. The use of frameworks such as React is strictly prohibited.

\hspace{2em}5. Only front‑end functionality needs to be implemented; no database or backend connections are required. If backend‑related functionality is involved, use mock data to simulate it.

\end{taskbox}

\subsection{Prompts for Evaluation Metrics}

Among all of our 24 evaluation metrics, 5 of them follow the LLM-as-a-judge paradigm. We show their prompts in Figures~\ref{Fig:generalFunctionPE} and~\ref{Fig:generalFunctionPECont} (General Functionality Correctness), Figures~\ref{Fig:generalVisualPE} and~\ref{Fig:generalVisualPECont} (General Visual Experience), Figure~\ref{Fig:functionAlignPE} (Functional Alignment), Figure~\ref{Fig:visualAlignPE} (Visual Alignment), and Figure~\ref{Fig:contentAlignPE} (Content Alignment), respectively.

\begin{figure*}[htbp]
    \begin{taskboxunbreak}[Prompts for the evaluation metric of General Functionality Correctness:]
\#\# [Role]

You are a senior front-end architect and testing lead, proficient in code review, white-box testing, front-end security, performance optimization, accessibility, and industry business standards.

\#\# [Current Time]

The current system time is \{date\}.

\#\# [Objective]

Your task is to evaluate the quality of the web page code and assign a score from 0 to 10 for each of the following 10 criteria to reflect its quality.

A score of 10 indicates the code is perfect, with no issues found during the code review.

A score of 0 indicates the code has severe syntax errors or major vulnerabilities, preventing it from running correctly.

A score from 1 to 9 indicates the code runs correctly, with higher scores representing better performance on the respective criterion.

Please output a comma-separated list of 10 numbers enclosed in square brackets, for example: [9,8,6,4,2,0,0,0,0].

Each scoring point is independent; please consider and score each one separately.

\#\#\# [Ten Evaluation Criteria and Scoring Guidance Examples]

1.  **Functional Completeness \& Business Logic**: Based on business requirements, ensure all functions are implemented without omission, the logic aligns with business specifications, and check that static data conforms to scientific and business common sense.

    **Scoring Guide**:
    
    \hspace{2em}**10 points**: All business functions are fully implemented, logic aligns with business requirements, and static data conforms to scientific common sense.
        
    \hspace{2em}**7-9 points**: Most functions are implemented, with minor flaws in the handling of a few features.
        
    \hspace{2em}**4-6 points**: Some business functions are incomplete, or there are errors in logic or issues with static data.
        
    \hspace{2em}**0-3 points**: Severe omissions in business functions, or logic is incorrect or does not meet business requirements.

2.  **Output Validation**: Following the code execution flow, evaluate if the output is correct. This includes checking value outputs and system calls, verifying for logical errors, missing, or duplicate output content. It's especially important to verify that UI updates and state changes reflect business logic changes.

    **Scoring Guide**:
        
    \hspace{2em}**10 points**: All outputs are as expected, data is correct, and UI and state updates are timely and complete.
        
    \hspace{2em}**7-9 points**: Most outputs are consistent with expectations, with minor inconsistencies in a few edge cases.
        
    \hspace{2em}**4-6 points**: There are inaccurate outputs or updates that do not occur as expected, potentially affecting the user experience.
        
    \hspace{2em}**0-3 points**: Outputs do not match expectations, system calls are not executed as required, affecting normal functionality.

3.  **Forms \& Critical Path Flows**: Includes pre-validation, disabled states, protection against duplicate submissions, success/failure notifications, and redirects. Ensures important flows like payments and bookings are idempotent, have state rollback mechanisms, and provide clear error messages, covering industry constraints (e.g., time windows, quantity limits).

    **Scoring Guide**:
        
    \hspace{2em}**10 points**: Form validation, disabled states, duplicate submission protection, and success/failure notifications are all complete. Critical paths like payments and bookings have robust idempotency and exception handling.
        
    \hspace{2em}**7-9 points**: Most form and critical path flows are handled well, but some minor details are imperfect.
        
    \hspace{2em}**4-6 points**: There are obvious flaws in form and critical path flows, leading to potential duplicate submissions or state management issues.
        
    \hspace{2em}**0-3 points**: Form and critical path flow handling is missing, severely impacting the normal progression of business processes.

4.  **Data Science Logic Validation**: Verify that static data and business logic within the code are reasonable, ensuring data conforms to scientific principles and industry standards. Check the accuracy of data processing methods, avoiding hard-coded values or illogical data assumptions.

    **Scoring Guide**:
    
    \hspace{2em}**10 points**: All static data is reasonable, and data processing methods align with industry standards and scientific common sense.
    
    \hspace{2em}**7-9 points**: Most data processing logic is reasonable, but some cases may have boundary issues or do not fully adhere to best practices.
    
    \hspace{2em}**4-6 points**: Data processing methods have errors or are unreasonable, potentially leading to business logic errors.
    
    \hspace{2em}**0-3 points**: Data processing methods are clearly unreasonable or conflict with industry common sense, affecting normal functionality.

5.  **List/Card Display**: Check the state management and interactive behavior of list and card components, ensuring that empty data placeholders and loading skeletons are implemented correctly, and error states are handled effectively with user notifications.

    **Scoring Guide**:
        
        \hspace{2em}**10 points**: List/card display is perfect. Empty data placeholders, loading skeletons, and error state retries all work correctly, providing an excellent user experience.
        
        \hspace{2em}**7-9 points**: Most display effects are good, but there are minor flaws in the display for certain states.
        
        \hspace{2em}**4-6 points**: Some display features are missing, or error states do not effectively notify the user.

    \end{taskboxunbreak}
\caption{Prompts for the evaluation metric of General Functionality Correctness.}
\label{Fig:generalFunctionPE}
\end{figure*}

\begin{figure*}[htbp]
    \begin{taskboxunbreak}[Prompts for the evaluation metric of General Functionality Correctness cont.:]
    \hspace{2em}**0-3 points**: Display is severely inadequate. Empty data placeholders, loading skeletons, and error state retries do not work, severely impacting the user experience.
    
6.  **Correctness \& Boundary Conditions**: Covers all boundary conditions, null/type checks, ensures resources are released correctly, avoids concurrency/race condition issues, and ensures functionality remains reliable under various extreme circumstances.

    **Scoring Guide**:
    
        \hspace{2em}**10 points**: The function performs perfectly under all boundary conditions, correctly handles null values and type checks, gracefully releases resources, and avoids concurrency/race conditions.
        
        \hspace{2em}**7-9 points**: Most boundary conditions are handled, but a few edge cases are not fully covered or have minor errors.
        
        \hspace{2em}**4-6 points**: The function fails to work correctly in some extreme cases, there are issues with resource release, or there are race conditions or omissions in null checks.
        
        \hspace{2em}**0-3 points**: Boundary conditions are not considered, there are multiple null or type errors, resources are not released properly, and race conditions are severe.

7.  **Security**: Includes input validation, output encoding, prevention of injection attacks (XSS/CSRF), and dependency risk control to ensure the code is free from common security vulnerabilities.

    **Scoring Guide**:
        
        \hspace{2em}**10 points**: Input validation is complete, output encoding is strict, prevention against injection/XSS/CSRF vulnerabilities is comprehensive, and dependency risks are fully controlled.
        
        \hspace{2em}**7-9 points**: Most security issues are addressed, but some input validation or dependencies have potential risks.
        
        \hspace{2em}**4-6 points**: Some security checks are missing, leaving potential vulnerabilities that could be exploited.
        
        \hspace{2em}**0-3 points**: Severe security vulnerabilities, such as XSS or CSRF attacks, are not prevented.

8.  **Branch \& State Coverage**: Ensures `if/else/switch/ternary` structures comprehensively cover critical paths and boundary cases, and handle early returns/exception branches; ensures proper management of variables, loading states, disabled states, error states, and empty states.

    **Scoring Guide**:
        
        \hspace{2em}**10 points**: All branch paths (including `if/else/switch`, etc.) are covered, early returns and exception handling are complete, and all states (loading, disabled, etc.) are managed reasonably.
        
        \hspace{2em}**7-9 points**: The vast majority of branches and states are covered, but a few paths or states are not fully handled.
        
        \hspace{2em}**4-6 points**: Some branches or states are not covered, which may lead to logical errors or unhandled exceptions.
        
        \hspace{2em}**0-3 points**: Critical branches are not covered, and state management is chaotic.

9.  **Data Consistency \& Flow Management**: Ensures DOM updates are consistent with the state, avoids race conditions or dirty data issues caused by global variables and closures, and reduces data flow conflicts.

    **Scoring Guide**:
        
        \hspace{2em}**10 points**: Data flow management is perfect, DOM updates are always consistent with the state, and there are no race conditions or dirty data issues from global variables/closures.
        
        \hspace{2em}**7-9 points**: Data flow is largely consistent, but there are a few minor inconsistencies or race condition issues.
        
        \hspace{2em}**4-6 points**: Data flow is poorly managed, with obvious race condition problems or dirty data risks.
        
        \hspace{2em}**0-3 points**: Data flow is severely chaotic, DOM updates are inconsistent with the state, and there are numerous race conditions and dirty data issues.

10. **Asynchronous Operations \& Error Handling**: `fetch/Promise/async` and other asynchronous operations have complete error handling, timeout control, and are designed with fallback mechanisms and user-friendly error messages.

    **Scoring Guide**:
        
        \hspace{2em}**10 points**: All asynchronous operations (`fetch/Promise/async`) handle exceptions and timeouts, and have complete fallback mechanisms and user-friendly error messages.
        
        \hspace{2em}**7-9 points**: Most asynchronous operations are handled well, but some exception or timeout handling is incomplete.
        
        \hspace{2em}**4-6 points**: Exception or timeout handling for asynchronous operations is inadequate, and error messages are unclear.
        
        \hspace{2em}**0-3 points**: Asynchronous operations do not handle exceptions or timeouts and lack error messages.

\#\# [Review Rule Requirements]

A score is required for each item.

\#\# [Output Rules]

Note: The final output should only contain the JSON content format. Do not wrap it in a Markdown JSON block.

\{\{

  \hspace{2em}"score": [1,2,3,4,5,6,7,8,9,10],
  
  \hspace{2em}"summary": [
  
    \hspace{2em}\{\{
    
      \hspace{4em}"evaluation1":"Evaluation Content",
      
      \hspace{4em}"score": "0-10",
      
      \hspace{4em}"reason": "Brief Reason"
      
    \hspace{2em}\}\}
  
  \hspace{2em}]
  
\}\}

user\_requirements: \{user\_requirements\}

Web Page Code: \{html\_content\}

    \end{taskboxunbreak}
\caption{Prompts for the evaluation metric of General Functionality Correctness continue.}
\label{Fig:generalFunctionPECont}
\end{figure*}

\begin{figure*}[htbp]
    \begin{taskboxunbreak}[Prompts for the evaluation metric of General Visual Experience:]
\# Role Setting
    
You are a senior product design reviewer with a keen aesthetic intuition and extensive user experience judgment. Please evaluate the first-screen interface of a front-end application from a real user's perspective, based on your subjective feelings.

\# Review Mindset

- Use first impressions as an important reference

- Trust your intuition and feelings

- View the application from the perspective of an ordinary user

- Don't get too caught up in technical details; focus on the "feel"

\# Subjective Evaluation Criteria

\#\# 0-1 Points - Design Lacking or Extremely Chaotic

- Almost no design awareness; the page appears extremely chaotic or incomplete

- Visual elements are piled up without order, lacking basic layout logic

- Color scheme is jarring or extremely disharmonious, causing strong discomfort

- Information is completely inaccessible; user experience is extremely poor

- Gives the impression of "this is a work in progress" or "something went wrong"

\#\# 1-2 Points - Basic Functionality Available, but Design is Rough

- Has basic information presentation capabilities, but severely lacks a sense of design

- Visual presentation is merely "viewable," lacking aesthetic appeal and refinement

- Uses standard templates or default styles with no signs of custom design

- Color scheme is mediocre or has obvious aesthetic issues (e.g., "tacky," "outdated")

- Layout is rigid, lacking visual hierarchy and breathing room

- Gives the feeling of "it's usable, but I don't want to use it"

\#\# 2-2.5 Points - Design is Acceptable, Conventional

- Design meets basic standards; visual presentation is relatively clean

- Color scheme is safe but lacks highlights, falling into the "not bad, but not great" category

- Layout is reasonable and information hierarchy is mostly clear, but lacks memorable features

- Uses common design patterns, giving a "deja vu" feeling

- Overall look and feel is ordinary; no obvious flaws, but fails to spark interest

- At a "passing grade" level; users won't dislike it, but won't be impressed either

\#\# 2.5-3 Points - Design is Good, with Clear Design Intent

- Has a clear design concept and visual style; overall cohesive and unified

- Color scheme is harmonious with a certain aesthetic pursuit, showing careful thought

- Layout is well-considered, information hierarchy is clear, and visual guidance is smooth

- Has highlights in certain details (e.g., animations, icons, typography)

- Style is relatively mature, matching the product's positioning and target users

- Overall quality is good, but innovation and distinctiveness are limited

- At a "good" level; users will find it professional and comfortable

\#\# 3-4 Points - Design is Excellent, Trend-setting

- Design style breaks through traditional paradigms, being both distinctive and forward-thinking, capable of leading trends in similar web design

- Theme is highly original, potentially incorporating unique cultures, niche areas, or innovative concepts to avoid homogenization

- Design philosophy is distinct, conveying clear brand values or core content through visual language, allowing users to perceive the unique design intent

- Content is deeply integrated with the design theme, supporting the visual presentation while being amplified by it, creating a synergistic "content-design" effect

- Visual presentation is refined and captivating, achieving a high degree of balance between aesthetics and functionality

- Evokes strong emotional resonance and is memorable, making users feel "wowed" or even compelled to "want to share"

\# Subjective Evaluation Dimensions

\#\# First Impression (Very Important!)

- The moment you open it, what is your gut reaction?

- Does it make you want to explore further, or close it?

- What is the overall "vibe"? Professional? Rough? Interesting? Boring?

\#\# Visual Experience (Use Your Aesthetics)

- Is it good-looking? Does it have lasting appeal?

- Is the color scheme harmonious? Does it feel "tacky"?

- Are there any "wow" factors?

- Is the overall feeling refined or rough?

- Is there a sense of design and quality?
    \end{taskboxunbreak}
\caption{Prompts for the evaluation metric of General Visual Experience.}
\label{Fig:generalVisualPE}
\end{figure*}

\begin{figure*}[htbp]
    \begin{taskboxunbreak}[Prompts for the evaluation metric of General Visual Experience cont.:]
\#\# Emotional Resonance

- Does this application have a "personality"?

- Does it feel warm? Cold? Professional? Friendly?

- Does it inspire a sense of trust?

\#\# Style Fit

- Does the visual style fit the product's industry? (e.g., finance should be stable, education friendly, e-commerce energetic)

- Does the design tone match the target user group? (e.g., trendy for young people, clear for the elderly, professional for business clients)

- Is there a sense of dissonance from a "style mismatch"? (e.g., using an overly playful design in a serious context)

\#\# Information Hierarchy

- Is the primary/secondary relationship on the page clear? Can you identify the main focus at a glance?

- Is the visual weight of titles, buttons, and supplementary information reasonable?

- Are important functions prominent enough? Is secondary information appropriately downplayed?

\#\# Design Consistency

- Is the visual expression of similar elements consistent? (e.g., button styles, icon styles)

- Is the color semantics consistent? (i.e., does the same color have a consistent meaning in different places)

- Are there any confusing design contradictions? (e.g., green indicating success in one place and in-progress in another)

\# Output Format

Strictly output in JSON format:

\{

  \hspace{2em}"first\_impression": "[Describe your feeling the moment you opened the app]",
  
  \hspace{2em}"overall\_summary": "[Summarize your overall impression of this app in one sentence]",
  
  \hspace{2em}"visual\_aesthetic\_evaluation": "[Subjective feelings on color scheme, refinement, sense of design, etc.]",
  
  \hspace{2em}"style\_fit": "[Whether the style fits the industry/user group, and if there's any dissonance]",
  
  \hspace{2em}"information\_hierarchy": "[Evaluation of primary/secondary relationships and prominence of key elements]",
  
  \hspace{2em}"design\_consistency": "[Evaluation of color semantics and element uniformity]",

  \hspace{2em}"if\_your\_friend\_made\_this": "[Provide feedback in more authentic and direct language]",
  
  \hspace{2em}"subjective\_score": [0-4 points, up to two decimal places],
  
  \hspace{2em}"scoring\_rationale": "[Explain the basis for your score, e.g., why it's X points and not X±0.3]"
  
\}

\# Review Philosophy

1. Trust your first intuition

2. Don't try to rationalize why you like or dislike it

3. React authentically like a regular user

4. There are no right answers in aesthetic judgment; trust your own feelings

5. Provide both a sentimental evaluation and a rational analysis of key elements like style, hierarchy, and consistency

6. Remain objective and friendly; when pointing out issues, offer direction rather than criticism

    \end{taskboxunbreak}
\caption{Prompts for the evaluation metric of General Visual Experience continue.}
\label{Fig:generalVisualPECont}
\end{figure*}

\begin{figure*}[htbp]
    \begin{taskboxunbreak}[Prompts for the evaluation metric of Functional Alignment:]

**Role-play:** You are a senior requirements evaluator, project manager, and user of web front-end applications. You have just received a web front-end application developed based on the given requirements. It is still in the early stages of development, and your task is to determine if the developer has correctly understood the user's needs and to judge at a high level whether the given web page code meets the user-specified requirements, without worrying about the correctness of the specific implementation.

**Task Goal:** Analyze the web front-end application code and determine if it meets the given requirements.

**Input:**

**Web page code** and **requirements** information.

**Evaluation Criteria:**

Please analyze whether the code implements the specified requirements from the user's perspective, based on the following dimensions:

\hspace{2em}**Functional Module:** Does the code contain the functionality specified in the requirement description? (Even if it's just a function to be implemented?)

\hspace{2em}**Interaction:** If the functionality involves user interaction (e.g., clicking a form submission button, an input box), is there a corresponding, user-visible control in the code, along with listener support?

**Output:**

Please output your judgment and analysis in JSON format, with the following structure:

[\{

\hspace{2em}"functional\_requirement": "<The requirement>",

\hspace{2em}"code\_snippet": "```<Web Front-end Application Code>```(The maximum output length is 1000 characters.)",

\hspace{2em}"is\_implemented": <true/false>,

\hspace{2em}"implementation\_analysis": "<A concise analysis of how the code implements the requirement>",

\hspace{2em}"confidence\_score": <A confidence score from 0 to 1, indicating your certainty in the judgment, where 1 is the highest>

\}]

**Example:**

**Input:**

1.  **Web page code:**

\hspace{2em}<button id="myButton">Click me</button>

\hspace{2em}<div id="message"></div>

\hspace{2em}<script>

\hspace{2em}\hspace{2em}document.getElementById("myButton").addEventListener("click", function() \{

\hspace{2em}\hspace{2em}document.getElementById("message").textContent = "The button was clicked!";

\hspace{2em}\hspace{2em}\});

\hspace{2em}</script>

2.  **Requirement(s):** [``After clicking a button, the message `The button was clicked!' will be displayed on the page.'']

**Expected Output (JSON format):**

Note: The final output should only contain the JSON content format, do not wrap it in a Markdown JSON block.

[\{

\hspace{2em}"functional\_requirement": "After clicking a button, the message 'The button was clicked!' is displayed on the page.",

\hspace{2em}"code\_snippet": 

\hspace{2em}\hspace{2em}"<button id="myButton">Click me</button>

\hspace{2em}\hspace{2em}<div id="message"></div>

\hspace{2em}\hspace{2em}<script>

\hspace{2em}\hspace{2em}\hspace{2em}document.getElementById("myButton").addEventListener("click", function() {

\hspace{2em}\hspace{2em}\hspace{2em}document.getElementById("message").textContent = "The button was clicked!";

\hspace{2em}\hspace{2em}\hspace{2em}});

\hspace{2em}\hspace{2em}</script>",

\hspace{2em}"is\_implemented": true,

\hspace{2em}"implementation\_analysis": "The code includes a button and a div element for displaying messages. JavaScript uses an event listener to bind a click event to the button. When the event is triggered, it modifies the textContent of the \#message element to 'The button was clicked!'. The functionality perfectly matches the description; clicking the button displays the specified message on the page in real-time.",

\hspace{2em}"confidence\_score": 1

\}]
    
    \end{taskboxunbreak}
\caption{Prompts for the evaluation metric of Functional Alignment.}
\label{Fig:functionAlignPE}
\end{figure*}

\begin{figure*}[htbp]
    \begin{taskboxunbreak}[Prompts for the evaluation metric of Visual Alignment:]
**Role-play:** You are a senior requirements reviewer, project manager, and a user of web front-end applications. You have just received a web front-end application developed according to given requirements. It is still in the early stages of development. Your task is to determine whether the developer has correctly understood the user requirements by evaluating at a high level whether the given web page code meets the specified requirement points, without needing to verify the correctness of the specific implementation.

**Task Goal:** Analyze the web front-end application code to determine if it meets the given requirement points.

**Input:**

**Web page code** and **requirement points** information

**Evaluation Criteria:**

Please analyze from the user's perspective whether the code implements the specified requirement points based on the following dimensions:

\hspace{2em}**Visual Attributes:** Does the code conform to the user-specified visual design?

\hspace{2em}**Page Components:** Does the code include the page components (must be user-visible) specified by the user?

**Output:**

Please output your evaluation results and analysis in JSON format, with the following structure:

[\{

\hspace{2em}"visual\_requirement": "<Requirement Point>",
  
\hspace{2em}"code\_snippet": "<Web Front-end Application Code> (The maximum output length is 1000 characters.)",
  
\hspace{2em}"is\_implemented": <true/false>,
  
\hspace{2em}"implementation\_analysis": "<A concise analysis of how the code implements the requirement point>",
  
\hspace{2em}"confidence\_score": <A confidence score from 0-1, indicating your certainty in the judgment, where 1 is the highest>
  
\}]

**Example:**

**Input:**

1.  **Web page code:**

\hspace{2em}.primary-button \{

\hspace{2em}background-color: \#007bff;

\hspace{2em}color: white;

\hspace{2em}font-size: 16px;

\hspace{2em}padding: 10px 20px;

\hspace{2em}border-radius: 5px;

\hspace{2em}\}

2.  **Requirement Points:** [``A button with a blue background, white text, a font size of 16 pixels, and a 5-pixel border-radius.'']

**Expected Output (JSON format):**

Note: The final output should only contain the JSON content, without being wrapped in Markdown's JSON format.

[\{

\hspace{2em}"visual\_requirement": "A button with a blue background, white text, a font size of 16 pixels, and a 5-pixel border-radius.",
  
\hspace{2em}"code\_snippet": 
  
\hspace{2em}\hspace{2em}".primary-button \{

\hspace{2em}\hspace{2em}background-color: \#007bff;

\hspace{2em}\hspace{2em}color: white;

\hspace{2em}\hspace{2em}font-size: 16px;

\hspace{2em}\hspace{2em}padding: 10px 20px;

\hspace{2em}\hspace{2em}border-radius: 5px;

\hspace{2em}\hspace{2em}\}",
  
\hspace{2em}"is\_implemented": true,
  
\hspace{2em}"implementation\_analysis": "The `.primary-button' class in the code sets the background color to \#007bff, which is a shade of blue, the text color to white, the font size to 16px, and the border-radius to 5px, fully matching the requirement's description. Additionally, the padding is set to 10px 20px, which is a common padding for buttons; although not mentioned in the requirement, it does not affect the consistency of the implementation.",
  
\hspace{2em}"confidence\_score": 1
  
\}]

    \end{taskboxunbreak}
\caption{Prompts for the evaluation metric of Visual Alignment.}
\label{Fig:visualAlignPE}
\end{figure*}

\begin{figure*}[htbp]
    \begin{taskboxunbreak}[Prompts for the evaluation metric of Content Alignment:]
**Role-play:** You are a senior requirements analyst, project manager, and a user of web front-end applications. You have just received a web front-end application developed based on given requirements. It is still in the early stages of development. Your task is to determine whether the developer has correctly understood the user's requirements and to judge at a high level whether the given web page code meets the user-specified requirement points, without worrying about the correctness of the specific implementation.

**Task Goal:** Analyze the web front-end application code and determine if it meets the given requirement points.

**Input:**

**Web page code** and **requirement points** information

**Evaluation Criteria:**

Please analyze from the user's perspective whether the code implements the specified requirement points based on the following dimensions:

\hspace{2em}**Content Semantics:** Does the content on the page conform to the semantics specified in the user requirements?

\hspace{2em}**Specified Data:** Does the content on the page include the specific data that the user specified to include? (e.g., image links, data source links, text copy, video/audio links, etc.)

**Output:**

Please output your judgment and analysis in JSON format, with the following structure:

[\{

\hspace{2em}"content\_requirement": "<Requirement Point>",
  
\hspace{2em}"code\_snippet": "<Web Front-end Application Code> (The maximum output length is 1000 characters.)",
  
\hspace{2em}"is\_implemented": <true/false>,
  
\hspace{2em}"implementation\_analysis": "<A concise analysis of how the code implements the requirement point>",
  
\hspace{2em}"confidence\_score": <A confidence score from 0-1, indicating your level of certainty, with 1 being the highest>

\}]

**Example:**

**Input:**

1.  **Code Snippet:**

\hspace{2em}<div>Welcome</div>
    
\hspace{2em}<img src="logo.png">
    
\hspace{2em}<a href="/home"></a>

2.  **Requirement Points:** ["Display a welcome text.", "Display a company logo, using logo.png.", "Provide a link to return to the homepage."]

**Expected Output (JSON format):**
Note: The final output should only contain the JSON content, do not wrap it in Markdown's JSON format.

[\{

\hspace{2em}"content\_requirement": "Display a welcome text.",

\hspace{2em}"code\_snippet": "<div>Welcome</div>",

\hspace{2em}"is\_implemented": true,

\hspace{2em}"implementation\_analysis": "The code uses a `<div>'tag to contain the text \"Welcome\", thus implementing the requirement to display a welcome text.",

\hspace{2em}"confidence\_score": 0.8

\},

\{

\hspace{2em}"content\_requirement": "Display a company logo.",

\hspace{2em}"code\_snippet": "<img src=\"logo.png\">",

\hspace{2em}"is\_implemented": true,

\hspace{2em}"implementation\_analysis": "The code uses an `<img>' tag to attempt to display the user-specified image named \"logo.png\".",

\hspace{2em}"confidence\_score": 1.0

\},

\{

\hspace{2em}"content\_requirement": "Provide a link to return to the homepage.",

\hspace{2em}"code\_snippet": "<a href=\"/home\"></a>",

\hspace{2em}"is\_implemented": false,

\hspace{2em}"implementation\_analysis": "The code uses an `<a>' tag to create a link pointing to \"/home\", but there is no text on the link, which fails to indicate to the user that it is a link to return to the homepage.",

\hspace{2em}"confidence\_score": 0.9

\}]

    \end{taskboxunbreak}
\caption{Prompts for the evaluation metric of Content Alignment.}
\label{Fig:contentAlignPE}
\end{figure*}

\end{document}